\documentclass[a4paper,UKenglish]{lipics-v2018}

\usepackage{url} 
\usepackage{algorithm}
\usepackage{algpseudocode}
\usepackage{amsmath}
\usepackage{graphicx}
\usepackage{bigstrut}
\usepackage{array}
\usepackage{listings}
\usepackage{enumitem}
\usepackage{color}
\usepackage{cleveref}
\usepackage{tikz}
\usepackage{semantic}
\usepackage{syntax,etoolbox}
\AtBeginEnvironment{grammar}{\footnotesize}
\usepackage[normalem]{ulem}
\useunder{\uline}{\ul}{}
\usepackage[utf8]{inputenc}
\usepackage{tikz}
\usepackage{sepnum}
\usepackage{adjustbox}
\usetikzlibrary{snakes,arrows,shapes}
\usepackage{inconsolata}
\usepackage{subcaption}
\usepackage{multicol}
\usepackage{etoolbox}

\crefformat{section}{\S#2#1#3}
\crefformat{subsection}{\S#2#1#3}

\hypersetup{pdftitle={},
  pdfsubject={}, 
  pdfauthor={},
  pdfkeywords={}, 
  pdfstartview=FitH,
  pdfpagemode={UseOutlines},
  bookmarksnumbered=true, bookmarksopen=true, colorlinks,
    citecolor=black,%
    filecolor=black,%
    linkcolor=black,%
    urlcolor=black}

\usepackage[all]{hypcap}

\definecolor{lightgray}{rgb}{0.95, 0.95, 0.95}
\definecolor{darkgray}{rgb}{0.4, 0.4, 0.4}
\definecolor{editorGray}{rgb}{0.95, 0.95, 0.95}
\definecolor{editorOcher}{rgb}{1, 0.5, 0} 
\definecolor{editorGreen}{rgb}{0, 0.5, 0} 
\definecolor{orange}{rgb}{1,0.45,0.13}
\definecolor{olive}{rgb}{0.17,0.59,0.20}
\definecolor{brown}{rgb}{0.69,0.31,0.31}
\definecolor{purple}{rgb}{0.38,0.18,0.81}
\definecolor{lightblue}{rgb}{0.1,0.57,0.7}
\definecolor{lightred}{rgb}{1,0.4,0.5}
\definecolor{editorRed}{RGB}{170, 8, 32}

\newcommand{\point}[1]{\par\smallskip\noindent{\bf #1:}}
\newcommand{\dde}[1]{\par\smallskip\emph{#1:}}

\algrenewcommand\algorithmicrequire{\textbf{Input:}}
\algrenewcommand\algorithmicensure{\textbf{Output:}}

\lstdefinelanguage{JavaScript}{
  morekeywords={typeof, new, true, false, throw, function, return, null, switch, default, var, if, in, while, do, else, case, break, instanceof},
  morecomment=[s]{/*}{*/},
  morecomment=[l]//,
  morestring=[b]",
  morestring=[b]'
}

\usepackage{float}
\usepackage{tcolorbox}
\newcommand{\nnum}[1]{\sepnum{.}{,}{}{#1}}

\newcommand{\empirical}[1]{{#1}}

\usepackage{microtype}

\title{Static Analysis for Asynchronous JavaScript Programs}

\titlerunning{Static Analysis for Asynchronous JavaScript Programs}

\author{Thodoris Sotiropoulos}{Athens University of Economics and Business, Greece}{theosotr@aueb.gr}{}{}
\author{Benjamin Livshits}{Imperial College London, UK}{b.livshits@imperial.ac.uk}{}{}
\authorrunning{ }

\subjclass{%
    \ccsdesc[500]{Theory of computation~Program analysis},
    \ccsdesc[300]{Software and its engineering~Semantics}
}%

\keywords{static analysis, asynchrony, JavaScript}

\category{}

\relatedversion{}

\supplement{}

\funding{}

\EventEditors{John Q. Open and Joan R. Access}
\EventNoEds{2}
\EventLongTitle{42nd Conference on Very Important Topics (CVIT 2016)}
\EventShortTitle{CVIT 2016}
\EventAcronym{CVIT}
\EventYear{2016}
\EventDate{December 24--27, 2016}
\EventLocation{Little Whinging, United Kingdom}
\EventLogo{}
\SeriesVolume{42}
\nolinenumbers

\hideLIPIcs  

\begin{document}
\maketitle

\begin{abstract}
  Asynchrony has become an inherent element of JavaScript,
  as an effort to improve the scalability
  and performance of modern web applications.
  To this end,
  JavaScript provides programmers
  with a wide range of constructs and features for
  developing code
  that performs asynchronous computations,
  including but not limited to timers,
  promises, and non-blocking I/O.
  
  However, the data flow imposed by asynchrony is implicit,
  and not always well-understood by the developers
  who introduce many asynchrony-related bugs to their programs.
  Worse,
  there are few tools
  and techniques available
  for analyzing and reasoning about such asynchronous applications.
  In this work, we address this issue by designing
  and implementing one of the first static analysis
  schemes capable of dealing with almost all
  the asynchronous primitives of JavaScript up to the 7th edition of
  the ECMAScript specification.
  
  Specifically, we introduce the \emph{callback graph},
  a representation for capturing data flow
  between asynchronous code.
  We exploit the callback graph for designing
  a more precise analysis
  that respects the execution order
  between different asynchronous functions.
  We parameterize our analysis
  with one novel context-sensitivity flavor,
  and we end up with multiple analysis variations
  for building callback graph.
  
  We performed a number of experiments
  on a set of hand-written
  and real-world JavaScript programs.
  Our results show that our analysis can be applied
  to medium-sized programs
  achieving~\empirical{79\%} precision on average.
  The findings further suggest
  that analysis sensitivity is beneficial
  for the vast majority of the benchmarks.
  Specifically, it is able to improve precision
  by up to~\empirical{28.5\%},
  while it achieves an~\empirical{88\%} precision
  on average without highly sacrificing performance.
  
\end{abstract}

\section{Introduction}

JavaScript is an integral part of web development. 
Since its initial release in 1995,
it has evolved from a simple scripting language---primarily
used for interacting with web pages---into
a complex and general-purpose programming language
used for developing both client- and server-side applications.
The emergence of Web 2.0 along with
the dynamic features of JavaScript,
which facilitate a flexible and rapid development,
have led to a dramatic increase in its popularity.
Indeed,
according to the annual statistics provided by Github,
which is the leading platform for hosting open-source software,
JavaScript is by far the most popular and active
programming language from 2014 to 2018~\cite{githubstats}.

Although the dominance of JavaScript is impressive,
the community has widely criticized it
because it poses many concerns as to the security
or correctness of the programs~\cite{dynstudy}.
JavaScript is a language
with a lot of dynamic and metaprogramming features,
including but not limited to
prototype-based inheritance,
dynamic property lookups,
implicit type coercions,
dynamic code loading, etc.
Many developers often do not understand
or do not properly use these features,
introducing errors to their programs---which
are difficult to debug---or
baleful security vulnerabilities.
In this context,
JavaScript has attracted many
engineers and researchers over the past decade to
1) study and reason about its peculiar characteristics,
and 2) develop new tools
and techniques---such as type analyzers~\cite{tajs,safe,jsai},
IDE and refactoring tools~\cite{refstatic,refstatic2,wala,async-refactoring},
or bug and vulnerability detectors~\cite{Maffeis2,gatekeeper,webracer,datamatter,fuzz,synode}---to
assist developers with
the development
and maintenance of their applications.
Program analysis,
and especially static analysis,
which automatically computes facts about
program's behavior without actually running it,
plays a crucial role in the design of such tools~\cite{jsanal}.

Additionally,
preserving the scalability of modern web applications
has become more critical than ever.
As an effort to improve the throughput of web programs,
JavaScript has started to adopt
an event-driven programming paradigm~\cite{event-driven}.
In this context,
a code is executed \emph{asynchronously}
in response to certain events,
e.g., user input,
a response from a server,
data read from disk, etc.
In the first years of JavaScript,
someone could come across that asynchrony
mainly in a browser environment
e.g., DOM events,
AJAX calls,
timers, etc.
However,
in recent years,
asynchrony has become a salient
and intrinsic element of the language,
as newer versions of the language's core specification
(i.e., ECMAScript) have introduced
more and more asynchrony-related features.
For example,
ECMAScript 6 introduces promises;
an essential element of asynchronous programming
that allows developers to track
the state of an asynchronous computation easily.
Specifically,
the state of a promise object can be one of:
\begin{itemize}
  \item \emph{fulfilled:} the associated operation is complete,
    and the promise object tracks its resulting value.
  \item \emph{rejected:} the associated operation failed,
    and the promise object tracks its erroneous value.
  \item \emph{pending:} the associated operation has been
    neither completed nor failed.
\end{itemize}
Promises are particularly useful for asynchronous programming
because they provide an intuitive way
for creating chains of
asynchronous computation,
facilitating the enforcement of execution order
as well as error propagation~\cite{callback-study,async-refactoring}.
To do that,
promises trigger the execution of certain functions
(i.e., \emph{callbacks}) depending on their state,
e.g., callbacks that are executed once a promise is fulfilled
or rejected.
For that reason,
the {\tt API} of promises provides the method {\tt x.then(f1, f2)}
for registering new callbacks (i.e., {\tt f1} and {\tt f2})
on a promise object {\tt x}.
For example,
we call the callback {\tt f1}
when the promise is fulfilled,
while we trigger the callback {\tt f2}
once the promise is rejected.
The method {\tt x.then()}
returns a new promise
which the return value of the provided callbacks (i.e., {\tt f1, f2}) fulfills.
Since their initial introduction to the language,
JavaScript developers
have widely embraced promises;
a study in 2015 showed that
75\% of JavaScript frameworks
use promises~\cite{callback-study}.

Building upon promises,
newer versions of ECMAScript have added
new language features related to asynchrony.
Specifically,
in ECMAScript 8,
we have the {\tt async/await} keywords.
The {\tt async} keyword declares a function
as asynchronous
which returns a promise fulfilled
with its return value,
while {\tt await x} defers the execution
of the \emph{asynchronous} function
in which is placed,
until the promise object {\tt x} is \emph{settled}
(i.e., it is either fulfilled or rejected).
The latest edition of ECMAScript
(ECMAScript 9)
adds asynchronous iterators and generators
that allow developers to iterate
over asynchronous data sources.

\begin{figure}[t]
\begin{lstlisting}
asyncRequest(url, options)
  .then(function (response) {
    honoka.response = response.clone();

    switch (options.dataType.toLowerCase()) {
      case "arraybuffer":
        return honoka.response.arrayBuffer();
      case "json":
        return honoka.response.json();
      ...
      default:
        return honoka.response.text();
    }
  })
  .then(function (responseData) {
    if (options.dataType === "" || options.dataType === "auto") {
      const contentType = honoka.response.headers.get("Content-Type");
      if (contentType && contentType.match("/application\/json/i")) {
        responseData = JSON.parse(responseData);
      }
    }
    ...
  });
\end{lstlisting}
\caption{Real-world example that mixes promises with asynchronous I/O.}\label{fig:motiv-ex-honoka}
\end{figure}

Beyond promises,
many JavaScript applications are written
to perform non-blocking I/O operations.
Unlike traditional statements,
when we perform a non-blocking I/O operation,
the execution is not interrupted
until that operation terminates.
For instance,
a file system operation is done asynchronously,
which means that the execution proceeds
to the next tasks
while I/O takes place.
Programmers often mix
asynchronous I/O with promises.
For instance,
consider the real-world example of Figure~\ref{fig:motiv-ex-honoka}.
At line 1,
the code performs an asynchronous request
and returns a promise object
which is fulfilled asynchronously
once the request succeeds.
Then that promise object
can be used for processing
the response of the server asynchronously.
For instance,
at lines 2--23,
we create a promise chain.
The first callback of this chain (lines 2--14)
clones the response of the request,
and assigns it to the property {\tt response}
of the object {\tt honoka} (line 3).
Then,
it parses the body of the response according to its type
and fulfills the promise object allocated by
the first invocation of {\tt then()}.
The second callback (lines 15--23)
retrieves the headers of the response---which
the statement at line 3 assigns to {\tt honoka.response}---and
if the content type is ``application/json'',
it converts the data of the response into a JSON object
(lines 17--19).

Like the other characteristics of JavaScript,
programmers do not always clearly understand asynchrony,
as a large number of asynchrony-related questions
issued in popular sites like
{\tt stackoverflow.com}\footnote{\url{https://stackoverflow.com/}}~\cite{madsen2015ev,promises},
or the number of bugs
reported in open-source repositories~\cite{concstudy,fuzz} indicate. 
However,
existing tools
and techniques have limited
(and in many cases no) support for asynchronous programs.
In particular,
existing tools
mainly focus on the event system of client-side
JavaScript applications~\cite{jensendom,parkdom},
and they lack the support of the more recent features
added to the language like promises.
Also,
many previous works conservatively considered
that all asynchronous callbacks
processed by the \emph{event loop}---the program point
which continuously waits for new events to come
and is responsible for the scheduling
and execution of callbacks---can
be called in any order~\cite{jensendom,parkdom,jsai}.
However,
such an approach may lead to imprecision
and false positives.
Back to the example of Figure~\ref{fig:motiv-ex-honoka},
it is easy to see
that an analysis,
which does not respect the execution order
between the first and the second callback,
will report a type error at line 17
(access of {\tt honoka.response.headers.get (" Content - Type ")}).
Specifically,
an imprecise analysis assumes
that the callback defined at lines 15--23
might be executed first;
therefore,
{\tt honoka.response},
assigned at line 3,
might be uninitialized.

In this work,
we tackle those issues,
by designing and implementing a static analysis
that deals with asynchronous JavaScript programs.
For that purpose,
we first define a model for understanding and
expressing a wide range of
JavaScript's asynchronous constructs,
and then we design a static analysis based on that.
We propose a new representation,
which we call \emph{callback graph},
which provides information
about the execution order of the asynchronous code.
The callback graph
proposed in this work
tries to shed light on
how data flow between asynchronous code is propagated.
Contrary to previous works,
we leverage the callback graph
and devise a more precise analysis
which respects the execution order of
asynchronous functions.
Furthermore,
we parameterize our analysis
with one novel context-sensitivity strategy
designed for asynchronous code.
Specifically,
we distinguish data flow
between asynchronous callbacks
based on the promise object on
which they have registered
or the next computation
to which execution proceeds.


\noindent
\point{Contributions} Our work makes the following four contributions:
\begin{itemize}
\item We propose a calculus,
i.e., $\lambda_{\tt q}$,
for modeling the asynchronous features
in the JavaScript language,
including
timers,
promises,
and asynchronous I/O operations.
Our calculus is a variation of
existing calculi~\cite{asyncjs,promises},
and provides constructs
and domains specifically targeted for our analysis
(\cref{sec:modeling}).

\item We design and implement a static analysis
that is capable of handling asynchronous JavaScript programs
by exploiting the abstract version of $\lambda_{\tt q}$.
To the best of our knowledge,
our analysis is the first to deal with JavaScript promises
(\cref{sec:tajs-domains}).

\item We propose the \emph{callback graph},
a representation
which illustrates the execution order
between asynchronous functions.
Building on that,
we propose a more precise analysis,
(i.e., \emph{callback-sensitive} analysis)
which internally consults the callback graph
to retrieve information about
the temporal relations of asynchronous functions
so that it propagates data flow accordingly.
Besides that,
we parameterize our analysis
with a novel context-sensitivity flavor
(i.e., \emph{QR-sensitivity})
used for distinguishing asynchronous callbacks.
(\cref{sec:callback-graph}, \cref{sec:analysis-sensitivity}).

\item We evaluate the performance
and precision of our analysis on a set of
micro benchmarks
and a set of real-world JavaScript modules.
For the impatient reader,
we find that our prototype is able to analyze
medium-sized asynchronous programs,
and the analysis sensitivity is beneficial
for improving the analysis precision.
The results showed
that our analysis is able to achieve
a~\empirical{79\%} precision for the callback graph,
on average.
The analysis sensitivity (i.e. callback- and QR-sensitivity)
can further improve callback graph precision
by up to~\empirical{28.5\%}
and reduce the total number of type errors
by~\empirical{16,7\%} as observed in the real-world benchmarks
(\cref{sec:evaluation}).

\end{itemize}


\section{Modeling Asynchrony}
\label{sec:modeling}

As a starting point,
we need to define a model
to express asynchrony.
The goal of this model is to provide us
with the foundations
for gaining a better understanding of the asynchronous primitives
and ease the design of a static analysis
for asynchronous JavaScript programs.
This model is expressed through
a calculus called $\lambda_{{\tt q}}$;
an extension of $\lambda_{{\tt js}}$
which is the core calculus
for JavaScript developed by~\cite{ljs}.
The $\lambda_{\tt q}$ calculus is designed to
be flexible so that
it can model various sources of asynchrony
found in the language
up to the 7th edition of ECMAScript.
(i.e., promises, timers, asynchronous I/O).
However,
it does not handle
the {\tt async/await} keywords
and the asynchronous iterators/generators
introduced in recent editions of the specification.

\subsection{The $\lambda_{\tt q}$ calculus}
\label{sec:calculus}

The key component of our model is \emph{queue objects}.
Queue objects are closely related to JavaScript promises.
Specifically,
a queue object---like a promise---tracks
the state of an asynchronous job,
and it can be in one of the following states:
1) \emph{pending},
2) \emph{fulfilled}
or 3) \emph{rejected}.
A queue object may trigger the execution of callbacks
depending on its state.
Initially,
a queue object is pending.
A pending queue object can transition
to a fulfilled or a rejected queue object.
A queue object might be fulfilled or rejected with
a provided value
which is later passed as an argument in
the execution of its callbacks.
Once a queue object is either fulfilled or rejected,
its state is final and cannot be changed.
We keep the same terminology as promises,
so if a queue object is either fulfilled or rejected,
we call it \emph{settled}.

\subsubsection{Syntax and Domains}

\begin{figure}[t]
  \begin{center}
  \begin{minipage}[5cm]{\linewidth}
  \begin{grammar}
    <$v \in Val$> ::= ...
      \alt $\bot$

    <$e \in Exp$> ::= ...
      \alt {\tt newQ()}
      \alt $e$.{\tt fulfill}$(e)$
      \alt $e$.{\tt reject}$(e)$
      \alt $e$.{\tt registerFul}$(e, e,\dots)$
      \alt $e$.{\tt registerRej}$(e, e,\dots)$
      \alt {\tt append}$(e)$
      \alt {\tt pop()}
      \alt $\bullet$

    <$E$> ::= ...
      \alt $E$.{\tt fullfill}$(e)$\: |\: $v$.{\tt fulfill}$(E)$
      \alt $E$.{\tt reject}$(e)$\: |\: $v$.{\tt reject}$(E)$
      \alt $E$.{\tt registerFul}$(e, e,\dots)$\: |\: $v$.{\tt registerFul}$(v, \dots, E, e,\dots)$
      \alt $E$.{\tt registerRej}$(e, e,\dots)$\: |\: $v$.{\tt registerRej}$(v, \dots, E, e,\dots)$
      \alt {\tt append}$(E)$

  \end{grammar}
  \end{minipage}
  \end{center}
  \caption{Syntax of $\lambda_{\tt q}$.}\label{fig:syntax}
\end{figure}

Figure~\ref{fig:syntax} illustrates the syntax of $\lambda_{\tt q}$.
For brevity,
we present only the new constructs added to the language.
Specifically,
we add eight new expressions:
\begin{itemize}
  \item {\tt newQ()}: This expression creates a fresh pending
    queue object with no callbacks associated with it.
  \item $e_1$.{\tt fulfill}$(e_2)$: This expression fulfills
    the receiver (i.e., the expression $e_1$) with
    the value of $e_2$.
  \item $e_1$.{\tt reject}$(e_2)$: This expression rejects
    the receiver (i.e., the expression $e_1$) with
    the value of $e_2$.
  \item $e_1$.{\tt registerFul}$(e_2, e_3,\dots)$:
    This expression registers the callback $e_2$ to the receiver.
    This callback is executed \emph{only}
    when the receiver is fulfilled.
    This expression also expects another queue object
    passed as the second argument,
    i.e., $e_3$.
    This queue object will be fulfilled
    with the return value of the callback $e_2$.
    This allows us to model chains of promises where
    a promise resolves with the return value of
    another promise's callback.
    This expression can also receive optional parameters
    (i.e., expressed through ``\dots'')
    with which $e_2$ is called
    if the queue object is fulfilled with $\bot$ value.
    We will justify later the intuition behind that.
  \item $e_1$.{\tt registerRej}$(e_2, e_3,\dots)$:
    The same as $e$.{\tt registerFul}$(\dots)$ but this time
    the given callback is executed once the receiver is rejected.
  \item {\tt append}($e$): This expression
    appends the queue object $e$ to the top
    of the current queue chain.
    As we shall see later,
    the top element of a queue chain corresponds
    to the queue object
    that is needed to be rejected
    if the execution encounters
    an uncaught exception.
  \item {\tt pop()}: This expression pops
    the top element of the current queue chain.
  \item The last expression $\bullet$ stands for the event loop.
\end{itemize}

Observe that we use evaluation contexts~\cite{felleisen2009semantics,ljs,madsen2015ev,asyncjs,promises}
to express how the evaluation of an expression proceeds.
The symbol $E$ denotes
which sub-expression is currently being evaluated.
For instance,
$E$.{\tt fulfill}$(e)$ describes
that we evaluate the receiver of {\tt fulfill},
whereas $v$.{\tt fulfill}$(E)$
implies that the receiver has been evaluated
to a value $v$,
and the evaluation now lies on the argument of {\tt fulfill}.
Beyond those expressions,
the $\lambda_{\tt q}$ calculus introduces a new value,
that is,
$\bot$.
This value differs from {\tt null}
and {\tt undefined}
because it expresses
the absence of value
and it does not correspond to any JavaScript value.

\begin{figure}[t]
  \centering
  \footnotesize
  \begin{align*}
    a \in Addr &= \{l_i \mid i \in \mathbb Z^{*}\} \cup \{l_{time}, l_{io}\} \\
    \pi \in Queue &= Addr \hookrightarrow QueueObject \\
    q \in QueueObject &= QueueState \times Callback^{*} \times Callback^{*} \times Addr \\
    s \in QueueState &= \{{\tt pending}\} \cup (\{{\tt fulfilled, rejected}\} \times Val) \\
    clb \in Callback &= Addr \times F \times Val^{*} \\
    \kappa \in ScheduledCallbacks &= Callback^{*} \\
    \kappa \in ScheduledTimerIO &= Callback^{*} \\
    \phi \in QueueChain &= Addr^{*}
  \end{align*}
  \caption{Concrete domains of $\lambda_{\tt q}$.}\label{fig:runtime}
\end{figure}

Figure~\ref{fig:runtime} presents the domains
introduced in the semantics of $\lambda_{\tt q}$.
In particular,
a queue is a partial map of addresses to queue objects.
The symbol $l_i$---where $i$ is a positive integer---indicates
an address.
Notice that the set of the addresses also
includes two special reserved addresses,
i.e., $l_{time}$, $l_{io}$.
We use these two addresses
to store the queue objects
responsible for keeping the state of
callbacks related to timers
and asynchronous I/O respectively
(Section~\ref{sec:async-primitives} explains how
we model those JavaScript features).
A queue object is described by its state---recall
that a queue object is either pending or
fulfilled and rejected with a value---a
sequence of callbacks
executed on fulfillment,
and a sequence of callbacks
called on rejection.
The last element of a queue object
is an address
which corresponds to another queue object
which is dependent on the current,
i.e., it is settled whenever
the current queue object is settled,
and with the same state.
We create such dependencies
when we settle a queue object
with another queue object.
In this case,
the receiver is dependent on
the queue object used as an argument.

Moving to the domains of callbacks,
we see that a callback consists of an address,
a function,
and a list of values
(i.e., arguments of the function).
Note that
the first component
denotes the address of the queue object
which the return value of the function
is going to fulfill.
In the list of callbacks
$\kappa \in ScheduledCallbacks$,
we keep the order in
which callbacks are scheduled.
Note that we maintain one more
list of callbacks
(i.e., $\tau \in ScheduledTimerIO$)
where we store the callbacks registered on
the queue objects
located at the addresses $l_{time}, l_{io}$.
We defer the discussion
about why we keep two separate lists
until Section~\ref{sec:model-event-loop}.

A queue chain $\phi \in QueueChain$
is a sequence of addresses.
In a queue chain,
we store the queue object
that we reject,
if there is an uncaught exception
in the current execution.
Specifically,
when we encounter an uncaught exception,
we inspect the top element of the queue chain,
and we reject it.
If the queue chain is empty,
we propagate the exception to the call stack as usual.

\subsubsection{Semantics}

\begin{figure}[]
\scriptsize
\centering
  \[
  \inference{
    e \hookrightarrow e'
  }
  {\pi,\: \phi,\: \kappa,\: \tau,\:  E[e] \rightarrow \pi',\: \phi',\: \kappa',\: \tau',\: E[e']}[[e-context]]
  \]\\
  \[
  \inference{
    fresh\: \alpha &
    \pi' = \pi[\alpha \mapsto ({\tt pending},\: [],\: [],\: \bot)]
  }
  {\pi,\: \phi,\: \kappa,\: \tau,\:  E[{\tt newQ}()] \rightarrow \pi',\: \phi,\: \kappa,\: \tau,\: E[\alpha]}[[newQ]]
  \]\\
  \[
  \inference{
    v \neq \bot & ({\tt pending},\: t,\: k, l) = \pi(p) & v \not\in dom(\pi) & t' = \langle(\alpha,\: f,\: [v],\: r) \mid (\alpha,\: f,\: a,\: r) \in t\rangle \\
    \kappa' = \kappa :: t' & \chi = ({\tt fulfilled}, v) & \pi' = \pi[p \mapsto (\chi, [], [], l)] & p \neq l_{time} \land p \neq l_{io}
  }
  {\pi,\: \phi,\: \kappa,\: \tau,\: E[p.{\tt fulfill}(v)] \rightarrow \pi',\: \phi,\: \kappa',\: \tau,\: E[l.{\tt fulfill}(v)]}[[fulfill-pending]]
  \]\\
  \[
  \inference{
    ({\tt pending},\: t,\: k, l) = \pi(p) & v \in dom(\pi) \\
    p(v) = ({\tt pending},\: t',\: k', \bot) & \pi' = \pi[v \mapsto ({\tt pending},\: t',\: k',\: p)]
  }
  {\pi,\: \phi,\: \kappa,\: \tau,\: E[p.{\tt fulfill}(v)] \rightarrow \pi',\: \phi,\: \kappa,\: \tau,\: E[{\tt undef}]}[[fulfill-pend-pend]]
  \]\\
  \[
  \inference{
    ({\tt pending},\: t,\: k, l) = \pi(p) & v \in dom(\pi) & p(v) = (({\tt fulfilled},\ v'),\: t',\: k', m)
  }
    {\pi,\: \phi,\: \kappa,\: \tau,\: E[p.{\tt fulfill}(v)] \rightarrow \pi,\: \phi,\: \kappa,\: \tau,\: E[p.{\tt fulfill}(v')]}[[fulfill-pend-ful]]
  \]\\
  \[
  \inference{
    v = \bot & ({\tt pending},\: t,\: k, l) = \pi(p) & \kappa' = \kappa :: t\\
    \chi = ({\tt fulfilled}, v) & \pi' = \pi[p \mapsto (\chi, [], [], l)]
  }
  {\pi,\: \phi,\: \kappa,\: \tau,\:  E[p.{\tt fulfill}(v)]\rightarrow \pi',\: \phi,\: \kappa',\: \tau,\: E[l.{\tt fulfill}(v)]}[[fulfill-pending-$\bot$]]
  \]\\
  \[
  \inference{
    \pi(p)\downarrow_1 \neq {\tt pending}
  }
  {\pi,\: \phi\:, \kappa,\: \tau,\:  E[p.{\tt fulfill}(v)]\rightarrow \pi,\: \phi,\: \kappa,\: \tau,\: E[{\tt undef}]}[[fulfill-settled]]
  \]\\
  \[
  \inference{
    ({\tt pending}\:, t\:, k, l) = \pi(p) & t' = t \cdot (p',\: f,\: [n_1, n_2,\dots,n_n],\: r) \\
    \pi' = \pi[p \mapsto ({\tt pending},\: t',\: k, l)]
  }
  {\pi,\: \phi,\: \kappa,\: \tau,\:  E[p.{\tt registerFul}(f, p', r, n_1, n_2,\dots,n_n)] \rightarrow \pi,'\: \phi,\: \kappa,\: \tau,\: E[{\tt undef}]}[[registerFul-pending]]
  \]\\
  \[
  \inference{
    p \neq l_{time} \land p \neq l_{io} & (s,\: t\:, k\:, \chi) = \pi(p) & s\downarrow_1 = {\tt fulfilled} \\
    s\downarrow_2 \neq \bot & \kappa' = \kappa \cdot (p',\: f,\: [s\downarrow_2],\: r)
  }
  {\pi,\: \phi,\: \kappa,\: \tau,\:  E[p.{\tt registerFul}(f, p', r, n_1, n_2,\dots,n_n)] \rightarrow \pi,'\: \phi,\: \kappa',\: \tau,\: E[{\tt undef}]}[[registerFul-fulfilled]]
  \] \\
  \[
  \inference{
    p \neq l_{time} \land p \neq l_{io} & (s,\: t\:, k\:, \chi) = \pi(p) & s\downarrow_1 = {\tt fulfilled} \\
    s\downarrow_2 = \bot & \kappa' = \kappa \cdot (p',\: f,\: [n_1, n_2,\dots,n_n],\: r)
  }
  {\pi,\: \phi,\: \kappa,\: \tau,\:  E[p.{\tt registerFul}(f, p', r, n_1, n_2,\dots,n_n)] \rightarrow \pi,'\: \phi,\: \kappa',\: \tau,\: E[{\tt undef}]}[[registerFul-fulfilled-$\bot$]]
  \] \\
  \[
  \inference{
    p = l_{time} \lor p = l_{io} & (s,\: t\:, k\:, \chi) = \pi(p) & s\downarrow_1 = {\tt fulfilled} \\
    s\downarrow_2 = \bot & \tau' = \tau \cdot (p',\: f,\: [n_1, n_2,\dots,n_n],\: r)
  }
  {\pi,\: \phi,\: \kappa,\: \tau,\:  E[p.{\tt registerFul}(f, p', r, n_1, n_2,\dots,n_n)] \rightarrow \pi,'\: \phi,\: \kappa,\: \tau',\: E[{\tt undef}]}[[registerFul-timer-io-$\bot$]]
  \] \\
  \[
  \inference{
    p \in dom(\pi) & \phi' = p \cdot \phi
  }
  {\pi,\: \phi,\: \kappa,\: \tau,\:  E[{\tt append}(p)] \rightarrow \pi,\: \phi',\: \kappa,\: \tau,\: E[{\tt undef}]}[[append]]
  \]
  \[
  \inference{
  }
  {\pi,\: p \cdot \phi,\: \kappa,\: \tau,\:  E[{\tt pop()}] \rightarrow \pi,\: \phi,\: \kappa,\: \tau,\: E[{\tt undef}]}[[pop]]
  \]
  \[
  \inference{
    \phi = p \cdot \phi'
  }
  {\pi,\: \phi,\: \kappa,\: \tau,\:  E[{\bf err}\ v] \rightarrow \pi,\: \phi',\: \kappa,\: \tau,\: E[p.{\tt reject}(v)]}[[error]]
  \]\\
  \caption{The semantics of $\lambda_{{\tt q}}$}\label{fig:semantics}
\end{figure}
\normalsize


Equipped with the appropriate definitions of the
syntax and domains,
in Figure~\ref{fig:semantics},
we present the small-step semantics of $\lambda_{\tt q}$
which is an adaptation of~\cite{asyncjs,promises}.
Note that
we demonstrate the most representative rules of
our semantics;
we omit some rules for brevity.
For what follows,
the binary operation denoted by the symbol $\cdot$ means
the addition of an element to a list,
the operation indicated by :: stands for list concatenation,
while $\downarrow_i$ means the projection of the $i^{th}$ element.

The rules of our semantics adopt the following form:
$$
\pi,\: \phi,\: \kappa,\: \tau,\:  E[e] \rightarrow \pi',\: \phi',\: \kappa',\: \tau',\:  E[e'] 
$$
That form expresses
that a given queue $\pi$,
a queue chain $\phi$,
two sequences of callbacks $\kappa$ and $\tau$,
and an expression $e$ in
the evaluation context $E$ lead to
a new queue $\pi'$,
a new queue chain $\phi'$,
two new sequences of callbacks $\kappa'$ and $\tau'$,
and a new expression $e'$
in the same evaluation context $E$,
assuming that the expression $e$ is reduced to $e'$
(i.e., $e \hookrightarrow e'$).
The {\tt [e-context]} rule describes this behavior.

The {\tt [newQ]} rule creates a new queue object and
adds it to the queue using a fresh address.
This new queue object is pending,
and it does not have any callbacks related to it.

The {\tt [fulfill-pending]} rule demonstrates the case
when we fulfill a pending queue object with the value $v$,
where $v \neq \bot$
and does \emph{not} correspond to any queue object.
In particular,
we first change the state of the receiver object
from ``pending'' to ``fulfilled''.
In turn,
we update the already registered callbacks (if any)
by setting the value $v$ as the only argument of them.
Then,
we add the updated callbacks to
the list of scheduled callbacks $\kappa$
(assuming that the receiver is neither $l_{time}$ nor $l_{io}$).
Also, observe
that the initial expression is reduced to
$l.{\tt fulfill}(v)$,
that is,
if there is a dependent queue object $l$,
we also fulfill that queue object with the same value $v$.

The {\tt [fulfill-pend-pend]}
describes the scenario
of fulfilling a pending queue object $p$
with another pending queue object $v$.
In this case,
we update the queue $\pi$
by making the queue object $p$
to be dependent on $v$.
This means
that we settle $p$
whenever we settle $v$ and in the same way.
Notice that
both $p$ and $v$ remain pending.

The {\tt [fulfill-pend-ful]} rule
demonstrates the case
when we try to fulfill a pending queue object $p$
with the fulfilled queue object $v$.
Then,
$p$ resolves with the value
with which $v$ is fulfilled.
This is expressed by the resulting expression
$p.{\tt fulfill}(v')$.

The {\tt [fulfill-pending-$\bot$]} rule captures the case
when we fulfill a queue object with a $\bot$ value.
This rule is the same with the {\tt [fulfill-pending]} rule,
however,
this time we do not update the arguments of
any callbacks registered on the queue object $p$.

The {\tt [fulfill-settled]} rule illustrates the case
when we try to fulfill a settled queue object.
Notice that this rule neither
affects the queue $\pi$
nor the lists of scheduled callbacks $\kappa$ and $\tau$.

The {\tt [registerFul-pending]} rule adds the provided callback $f$
to the list of callbacks
that we should execute
once the queue object $p$ is fulfilled.
Note that this rule also associates this callback
with the queue object $p'$ given as the second argument.
That queue object $p'$ is fulfilled upon the termination of $f$.
Also,
this rule adds any extra arguments passed in {\tt registerFul}
as arguments of $f$.

The {\tt [registerFul-fulfilled]} rule adds
the given callback $f$ to the list  $\kappa$
(assuming that the receiver is neither $l_{time}$ nor $l_{io}$).
We use the fulfilled value of the receiver
as the only argument of the given function.
Like the previous rule,
it relates the provided queue object $p'$
with the execution of the callback.
This time we do ignore any extra arguments
passed in {\tt registerFul},
as we fulfill the queue object $p$ with
a value that is not $\bot$.

The {\tt [registerFul-fulfilled-$\bot$]} rule describes the case
where we register a callback $f$ on a queue object
fulfilled with a $\bot$ value.
Unlike the {\tt [registerFul-fulfilled]} rule,
this rule does not neglect
any extra arguments passed in {\tt registerFul}.
In particular,
it sets those arguments as parameters of the given callback.
This distinction allows us to pass
arguments explicitly to a callback.
Most notably,
these arguments are not dependent on the value
with which a queue object is fulfilled or rejected.

The {\tt [registerFul-timer-io-$\bot$]} rule is the same
as the previous one,
but this time we deal with queue objects
located either at $l_{time}$ or at $l_{io}$.
Thus,
we add the given callback $f$ to
the list $\tau$ instead of $\kappa$.

The {\tt [append]} rule appends the element
$p$ to the front of the current queue chain.
Note that this rule requires the element $p$
to be a queue object
(i.e., $p \in dom(\pi)$).
On the other hand,
the {\tt [pop]} rule removes the top element
of the queue chain.

The {\tt [error]} rule demonstrates the case
when we encounter an uncaught exception,
and the queue chain is not empty.
In that case,
we do not propagate the exception to the caller,
but we pop the queue chain
and get the top element.
In turn,
we reject the queue object $p$
specified in that top element.
In this way,
we capture the actual behavior of the uncaught exceptions
triggered during the execution of
an asynchronous callback.


\subsubsection{Modeling the Event Loop}
\label{sec:model-event-loop}

\begin{figure}[]
\scriptsize
\centering
  \[
  \inference{
    \kappa = (q, f, a) \cdot \kappa' & \phi = [] & \phi' = q \cdot \phi
  }
  {\pi\:, \phi,\: \kappa,\: \tau,\:  E[\bullet] \rightarrow \pi,\: \phi',\: \kappa',\: \tau,\: q{\tt.fulfill}(E[f(a)]); {\tt pop}(); \bullet}[[event-loop]]
  \] \\
  \[
  \inference{
    {\tt pick\: } (q, f, a)\: {\tt from\ } \tau & \tau' = \langle\rho\ \mid \forall \rho \in \tau.\: \rho \neq (q, f, a)\rangle\\
    \phi = [] & \phi' = q \cdot \phi
  }
  {\pi\:, \phi,\: [],\: E[\bullet] \rightarrow \pi,\: \phi',\: [],\: \tau',\: q{\tt.fulfill}(E[f(a)]); {\tt pop}(); \bullet}[[event-loop-timers-io]]
  \]
  \caption{{The semantics of the event loop.}}\label{fig:semantics-event-loop}
\normalsize
\end{figure}

A reader might wonder
why do we keep two separate lists,
i.e., the list $\tau$ for holding callbacks
coming from the $l_{time}$ or $l_{io}$ queue objects,
and the list $\kappa$ for callbacks of any other queue objects.
The intuition behind this design choice is that
it is convenient for us to
model the concrete semantics of the event loop correctly.
In particular,
the implementation of the event loop
assigns different priorities to the callbacks
depending on their kind~\cite{asyncjs,nodejs-schedule}.
For example,
the event loop processes
a callback of a promise object
before any callback of a timer
or an asynchronous I/O operation
regardless of their registration order.

In this context,
Figure~\ref{fig:semantics-event-loop} demonstrates the
semantics of the event loop.
The {\tt [event-loop]} rule pops
the first scheduled callback from the list $\kappa$.
Then,
we get the queue object
specified in that callback,
and we attach it to the front of the queue chain.
Adding the queue object $q$ to the top of
the queue chain allows us to reject that queue object
if there is an uncaught exception
during the execution of $f$.
In this case,
the evaluation of {\tt fulfill} will not have any
effect on the already rejected queue object $q$
(recall the {\tt [fullfill-settled]} rule).
Also,
observe how the event loop is reduced,
i.e., $q.{\tt fulfill}(f(a)); {\tt pop}(); \bullet$.
Specifically,
once we execute the callback $f$
and fulfill the dependent queue object $q$
with the return value of that callback,
we evaluate {\tt pop()},
that is,
we pop the top element of the queue chain
before we re-evaluate the event loop.
This is an invariant of the semantics of the event loop:
every time we evaluate it,
the queue chain is always empty.

The {\tt [event-loop-timers-io]} rule handles
the case when the list $\kappa$ is empty.
In other words,
that rule states
that if there are not any callbacks,
which neither come from the $l_{time}$ nor
the $l_{io}$ queue object,
inspect the list $\tau$,
and pick \emph{non-deterministically} one of those.
Selecting a callback non-deterministically
allows us to over-approximate the actual behavior
of the event loop regarding
its different execution phases~\cite{asyncjs}.
Overall,
that rule describes
the scheduling policy presented in
the work of~\cite{asyncjs},
where initially
we look for any callbacks of promises,
and if there exist,
we select one of those.
Otherwise,
we choose any callback associated
with timers and asynchronous I/O at random.

\subsubsection{Modeling Timers \& Asynchronous I/O}
\label{sec:model-timers-io}

\begin{figure}[t]
\small
\begin{center}
\begin{minipage}[5cm]{\linewidth}
\begin{grammar}
    <$e \in Exp$> ::= ...
      \alt {\tt addTimerCallback}$(e_1, e_2, e_3,\dots)$
      \alt {\tt addIOCallback}$(e_1, e_2, e_3,\dots)$

    <$E$> ::= ..
      \alt {\tt addTimerCallbackCallback}$(E, e,\dots)$\: |\: {\tt addTimerCallback}$(v, \dots, E, e,\dots)$
      \alt {\tt addIOCallback}$(E, e,\dots)$\: |\: {\tt addIOCallback}$(v, \dots, E, e,\dots)$
\end{grammar}
\end{minipage}
\end{center}
\caption{Extending the syntax of $\lambda_{\tt q}$ to deal with
timers and asynchronous I/O.}\label{fig:syntax-extended}
\end{figure}

\begin{figure}[h]
\scriptsize
\[
  \inference{
    q = \pi(l_{time})
  }
  {\pi,\: \phi,\: \kappa,\: \tau,\:  {\tt addTimerCallback}(f,\ n_1,\dots) \rightarrow \pi,\: \phi,\: \kappa,\: q.{\tt registerFul}(f,\ q,\ n_1,\dots)}[[add-timer-callback]]
\] \\
\[
  \inference{
    q = \pi(l_{io})
  }
  {\pi,\: \phi,\: \kappa,\: \tau,\:  {\tt addIOCallback}(f,\ n_1,\dots) \rightarrow \pi,\: \phi,\: \kappa,\: q.{\tt registerFul}(f,\ q,\ n_1,\dots)}[[add-io-callback]]
\]
\caption{Extending the semantics of $\lambda_{\tt q}$ to deal with
timers and asynchronous I/O.}\label{fig:semantics-extended}
\end{figure}

To model timers and asynchronous I/O,
we follow a similar approach to
the work of~\cite{asyncjs}.
Specifically,
we start with an initial queue $\pi$,
which contains two queue objects:
the $q_{time}$,
and $q_{io}$
which are located at
the $l_{time}$
and $l_{io}$ respectively.
Both $q_{time}$ and $q_{io}$
are initialized as
$(({\tt fulfilled},\: \bot),\: [],\: [],\: \bot)$.
Besides that,
we extend the syntax of $\lambda_{\tt q}$
by adding two more expressions.
Figure~\ref{fig:syntax-extended}
shows the extended syntax of $\lambda_{\tt q}$
to deal with timers and asynchronous I/O,
while Figure~\ref{fig:semantics-extended}
presents the rules related to those expressions.

The new expressions have high correspondence to each other.
Specifically,
the {\tt addTimerCallback($\dots$)} construct
adds the callback $e_1$
to the queue object located at the address $l_{time}$.
We call the provided callback
The arguments of that callback
are any optional parameters
passed in {\tt addTimerCallback},
i.e.,
$e_2, e_3$,
and so on.
From Figure~\ref{fig:semantics-extended},
we observe that
the {\tt [add-timer-callback]} rule retrieves
the queue object $q$ corresponding to
the address $l_{time}$.
Recall again that the $l_{time}$ can be found
in the initial queue.
Then,
the given expression is reduced to
$q.{\tt registerFul}(f,\ q,\ n_1,\dots)$.
In particular,
we add a new callback $f$ to the queue object
found at $l_{time}$.
Observe that we pass the same queue object
(i.e., $q$)
as the second argument of {\tt registerFul}.
That means that the execution of $f$
does not affect any queue object since
$q$ is already settled.
Recall that
according to the {\tt [fulfill-settled]} rule (Figure~\ref{fig:semantics}),
trying to fulfill (and similarly to reject)
a settled queue object does not have any effect.
Beyond that,
since $q$ is fulfilled with $\bot$,
the extra arguments (i.e., $n_1$,\dots)
are also passed as arguments in the invocation of $f$.

The semantics of the {\tt addIOCallback($\dots$)} primitive
is the same with that of {\tt addTimerCallback($\dots$)};
however,
this time,
we use the queue object located at $l_{io}$.

\subsection{Expressing Promises in Terms of $\lambda_{\tt q}$}
\label{sec:async-primitives}

The queue objects and their operations introduced in
$\lambda_{\tt q}$ are very closely related to
JavaScript promises.
Therefore,
the translation of promises' operations
into $\lambda_{\tt q}$ is straightforward.
We model every property and
method (except for {\tt Promise.all()})
by faithfully following the ECMAScript specification.

\begin{figure}[h]
\centering
\begin{lstlisting}[escapeinside={(*}{*)}]
Promise.resolve = function(value) {
  var promise = newQ();
  if (typeof value.then === "function") {
    var t = newQ();
    t.fulfill((*$\bot$*));
    t.registerFul(value.then, t, promise.fulfill, promise.reject);
  } else
    promise.fulfill(value);
  return promise;
}
\end{lstlisting}
\caption{
    Expressing {\tt Promise.resolve}
    in terms of $\lambda_{\tt q}$}\label{fig:modeling-resolve}.
\end{figure}

{\bf Example---Modeling Promise.resolve()}:
In Figure~\ref{fig:modeling-resolve},
we see how we model the {\tt Promise.resolve()} function
in terms of $\lambda_{\tt q}$.
The JavaScript {\tt Promise.resolve()} function creates
a new promise,
and resolves it with the given value.
According to ECMAScript,
if the given {\tt value} is a \emph{thenable},
(i.e., an object
which has a property named ``then''
and that property is a callable),
the created promise resolves asynchronously.
Specifically,
we execute the function {\tt value.then()} asynchronously,
and we pass the resolving functions
(i.e., {\tt fulfill, reject})
as its arguments.
Observe how the expressiveness of $\lambda_{\tt q}$
can model this source of asynchrony (lines 4--6).
First,
we create a fresh queue object {\tt t},
and we fulfill it with $\bot$ (lines 4, 5).
Then,
at line 6,
we schedule the execution of {\tt value.then()}
by registering it on the newly created queue object {\tt t}.
Notice that we also pass {\tt promise.fulfill}
and {\tt promise.reject} as extra arguments.
That means that those functions
will be the actual arguments of {\tt value.then()}
because {\tt t} is fulfilled with $\bot$.
On the other hand,
if {\tt value} is not a thenable,
we synchronously resolve the created promise
using the {\tt promise.fulfill} construct at line 8.

\section{The Core Analysis}
\label{sec:analysis}

The $\lambda_{\tt q}$ calculus presented
in Section~\ref{sec:modeling}
is the touchstone of the static analysis proposed
for asynchronous JavaScript programs.
The analysis is designed to be sound;
thus,
we devise abstract domains
and semantics
that over-approximate
the behavior of $\lambda_{\tt q}$.
Currently,
there are few implementations available
for asynchronous JavaScript,
and previous efforts mainly focus on
modeling the event system
of client-side applications~\cite{jensendom,parkdom}.
To the best of our knowledge,
it is the first static analysis for ES6 promises.
The rest of this section describes the details of the analysis.

\subsection{The Analysis Domains}
\label{sec:tajs-domains}

\begin{figure}[th]
  \footnotesize
  \centering
  \begin{align*}
      l \in  \widehat{Addr} &= \{l_i \mid i\ \text{is\ an\ allocation\ site}\} \cup \{l_{time}, l_{io}\} \\
      \pi \in \widehat{Queue} &=  \widehat{Addr} \hookrightarrow \mathcal{P}(\widehat{QueueObject})\\
      q \in \widehat{QueueObject} &= \widehat{QueueState} \times \mathcal{P}(\widehat{Callback}) \times \mathcal{P}(\widehat{Callback}) \times \mathcal{P}(\widehat{Addr}) \\
      qs \in \widehat{QueueState} &= \{{\tt pending}\} \cup (\{{\tt fulfilled, rejected}\} \times Value)\\
      clb \in \widehat{Callback} &= \widehat{Addr} \times \widehat{Addr} \times F \times Value^{*} \\
      \kappa \in \widehat{ScheduledCallbacks} &= (\mathcal{P}(\widehat{Callback}))^{*} \\
      \tau \in \widehat{ScheduledTimerIO} &= (\mathcal{P}(\widehat{Callback}))^{*} \\
      \phi \in \widehat{QueueChain} &= (\mathcal{P}(\widehat{Addr}))^{*} \\
  \end{align*}
  \caption{{The abstract domains of $\lambda_{\tt q}$}.}\label{fig:tajs-async-domains}
\end{figure}

Figure~\ref{fig:tajs-async-domains}
presents the abstract
domains of the $\lambda_{\tt q}$ calculus
that underpin our static analysis.
Below we make a summary of our primary design choices.

\dde{Abstract Addresses}
As a starting point,
we employ allocation site abstraction
for modeling the space of addresses.
It is the standard way used in literature
for abstracting addresses
which keeps the domain finite~\cite{tajs,madsen2015ev}.
Also note that
we still define two internal addresses,
i.e., $l_{time}, l_{io}$,
corresponding to the addresses of the
queue objects responsible for timers
and asynchronous I/O respectively.

\dde{Abstract Queue}
We define an abstract queue as
the partial map of abstract addresses to
an element of the power set of abstract queue objects.
Therefore,
an address might point to multiple queue objects.
This abstraction over-approximates
the behavior of $\lambda_{\tt q}$
and allows us to capture all possible program behaviors
that might stem from the analysis imprecision.

\dde{Abstract Queue Objects}
A tuple
consisting of an abstract queue state---observe
that the domain of abstract queue states
is the same as $\lambda_{\tt q}$---two
sets of abstract callbacks
(executed on fulfillment and rejection respectively),
and a set of abstract addresses
(used to store the queue objects
that are dependent on the current one)
represents an abstract queue object.
Notice how this definition differs from
that of $\lambda_{\tt q}$.
First,
we do not keep the registration order
of callbacks;
therefore,
we convert the two lists into two sets.
The programming pattern related to
promises supports our design decision.
Specifically,
developers often use promises as a chain;
registering two callbacks
on the same promise object is quite uncommon.
Madsen et. al.~\cite{madsen2015ev}
made similar observations for
the event-driven programs.

That abstraction can negatively affect precision
only when we register multiple callbacks on
a \emph{pending} queue object.
Recall from Figure~\ref{fig:semantics},
when we register a callback on a settled queue object,
we can precisely track its execution order
since we directly add it to the list of scheduled callbacks.
Second,
we define the last component as
a set of address;
something that enables us to
track all possible dependent queue objects soundly.

\dde{Abstract Callback}
An abstract callback comprises
two abstract addresses,
one function,
and a list of values
which stands for the arguments
of the function.
Note that the first address corresponds to
{\tt this} object,
while the second one is the queue object
which the return value of the function fulfills.

\dde{Abstract List of Scheduled Callbacks}
We use a list of sets
to abstract the domain responsible
for maintaining the callbacks
which are ready for execution
(i.e., $\widehat{ScheduledCallbacks}$
and $\widehat{ScheduledTimerIO}$).
In this context,
the $i^{th}$ element of a list
denotes the set of callbacks
that are executed after those placed
at the $(i-1)^{th}$ position
and before the callbacks located
at the $(i+1)^{th}$ position of the lists.
The execution of callbacks
of the same set is not known to the analysis;
they can be called in any order.
For example,
consider the following sequence $[\{x\}, \{y,\ z\}, \{w\}]$,
where $x, y, z, w \in \widehat{Callback}$.
We presume that the execution of elements
$y, z$ succeeds that of $x$,
and precedes that of $w$,
but we cannot compare $y$ with $z$,
since they are elements of the same set;
thus,
we might execute $y$ before $z$
and vice versa.

Note that a critical requirement of our domains' definition
is that they should be finite
so that the analysis is guaranteed to terminate.
Keeping the lists of scheduled callbacks bound is tricky
because the event loop might process
the same callback multiple times,
and therefore,
we have to add it to the lists $\kappa$ or $\tau$
more than one time.
For that reason,
those lists monitor the execution order of callbacks
up to a certain limit $n$.
The execution order of the callbacks
scheduled after that limit is not preserved;
thus,
the analysis places them in the same set.

\dde{Abstract Queue Chain}
The analysis uses
the last component of our abstract domains
to capture the effects of uncaught exceptions
during the execution of callbacks.
We define it as a sequence of sets of addresses.
Based on the abstract translation of
the semantics of $\lambda_{\tt q}$,
when the analysis reaches an uncaught exception,
it inspects the top element of
the abstract queue chain
and rejects all the queue objects found in that element.
If the abstract queue chain is empty,
the analysis propagates that exception
to the caller function as usual.
Note that the queue chain is
guaranteed to be bound.
In particular,
during the execution of a callback,
the size of the abstract queue chain is always one
because the event loop executes only one callback at a time.
The only case
when the size of the abstract queue chain is greater than 1 is
when we have nested promise executors.
A promise executor is a function passed as an argument
in a promise constructor.
However,
since we cannot have an unbound number of nested
promise executors,
the size of the abstract queue chain remains finite.

\subsubsection{Tracking the Execution Order}

{\bf Promises.} Estimating the order
in which
the event loop executes callbacks of promises
is straightforward
because it is a direct translation
of the corresponding semantics of $\lambda_{\tt q}$.
In particular,
there are two possible cases:

\begin{itemize}
    \item \emph{Settle a promise which has registered callbacks:}
        When we settle (i.e., either fulfill or reject) a promise object
        which has registered callbacks,
        we schedule those callbacks associated with the next
        state of the promise
        by putting them
        on the tail of the list $\kappa$.
        For instance,
        if we fulfill a promise,
        we append all the callbacks
        triggered on fulfillment on the list $\kappa$.
        A reader might observe
        that if there are multiple callbacks registered on
        the same promise object,
        we put them on the same set
        which is the element that we finally add to $\kappa$.
        The reason for this is that
        an abstract queue object does not
        keep the registration order of callbacks.
    \item \emph{Register a callback on an already settled promise:}
        When we encounter a statement of the form
        {\tt x.then(f1, f2)},
        where {\tt x} is a settled promise,
        we schedule either callback {\tt f1} or {\tt f2}
        (i.e., we add it to the list $\kappa$)
        depending on the state of that promise,
        i.e., we schedule callback {\tt f1}
        if {\tt x} is fulfilled
        and {\tt f2} if it is rejected.
\end{itemize}

\noindent
{\bf Timers \& Asynchronous I/O.} A static analysis
is not able to reason about the external environment.
For instance,
it cannot decide
when an operation on a file system
or a request to a server is complete.
Similarly,
it is not able to deal with time.
For that purpose,
we adopt a conservative approach
for tracking the execution order
between callbacks related to
timers and asynchronous I/O.
In particular,
we assume that the execution
order between those callbacks is unspecified;
thus,
the event loop might process them in any order.
However,
we \emph{do} keep track the execution order
between nested callbacks.

\subsection{Callback Graph}
\label{sec:callback-graph}

In this section,
we introduce the concept of \emph{callback graph};
a fundamental component of our analysis
that captures how data flow is propagated
between different asynchronous callbacks.
A callback graph is defined as an element of
the following power set:

$$
cg \in CallbackGraph = \mathcal{P}(Node \times Node)
$$
We define every node of a callback graph as
$n \in Node = C \times F$,
where $C$ is the domain of contexts
while $F$ is the set of all the functions of the program.
Every element of a callback graph $(c_1, f_1, c_2, f_2) \in cg$,
where $cg \in CallbackGraph$ has the following meaning:
\emph{the function $f_2$ in context $c_2$ is executed
immediately after the function $f_1$ in context $c_1$}.
We can treat the above statement as
the following expression: $f_1();f_2();$

\begin{definition}\label{def:1}
  Given a callback graph $cg \in CallbackGraph$,
  we define the binary relation $\rightarrow_{cg}$
  on nodes of the callback graph $n_1, n_2 \in Node$ as:
  $$
  n_1 \rightarrow_{cg} n_2 => (n_1, n_2) \in cg
  $$
\end{definition}

\begin{definition}\label{def:2}
  Given a callback graph $cg \in CallbackGraph$,
  we define the binary relation $\rightarrow_{cg}^{*}$
  on nodes of the callback graph $n_1, n_2 \in Node$ as:
  \begin{align*}
    n_1 \rightarrow_{cg} n_2 &=> n_1 \rightarrow_{cg}^{*} n_2\\
    n_1 \rightarrow_{cg}^{*} n_2 \land n_2 \rightarrow_{cg}^{*} n_3 &=> n_1 \rightarrow_{cg}^{*} n_3, \quad \text{where}\ n_3 \in Node
  \end{align*}
\end{definition}

Definition~\ref{def:1}
and Definition~\ref{def:2}
introduce the concept of a \emph{path}
between two nodes in a callback graph $cg \in CallbackGraph$.
In particular,
the relation $\rightarrow_{cg}$
denotes that there is path of length one between two nodes
$n_1, n_2$,
i.e., $(n_1, n_2) \in cg$.
On the other hand,
the relation $\rightarrow_{cg}^{*}$
describes
that there is a path of unknown length
between two nodes.
Relation $\rightarrow_{cg}^{*}$
is very important
as it allows us to identify the \emph{happens-before}
relation between two nodes $n_1, n_2$
even if $n_2$ is executed long after $n_1$,
that is $(n_1, n_2) \not\in cg$.
A significant property of a callback graph
is that it does \emph{not} have any cycles,
i.e.,

$$\forall n_1, n_2 \in Node.\: n_1 \rightarrow_{cg}^{*} n_2 => n_2 \not\rightarrow_{cg}^{*} n_1$$

Notice
that if $n_1 \not\rightarrow_{cg}^{*} n_2$,
and $n_2 \not\rightarrow_{cg}^{*} n_1$ hold,
the analysis cannot estimate
the execution order between $n_1$
and $n_2$.
Therefore,
we presume that $n_1$
and $n_2$ can be called in any order.

\subsection{Analysis Sensitivity}
\label{sec:analysis-sensitivity}

\subsubsection{Callback Sensitivity}

Knowing the temporal relations between asynchronous callbacks
enables us to capture how data flow is propagated precisely.
Typically,
a naive flow-sensitive analysis,
which exploits the control flow graph (CFG),
represents the event loop
as a single program point with only one context
corresponding to it.
Therefore---unlike traditional function calls---the
analysis misses
the happens-before relations between callbacks
because they are triggered by the same program location
(i.e., the event loop).

To address those issues,
we exploit the callback graph to devise
a more precise analysis,
which we call \emph{callback-sensitive} analysis.
The callback-sensitive analysis
propagates the state with regards to
the $\rightarrow_{cg}$
and $\rightarrow_{cg}^{*}$ relations
found in a callback graph $cg \in CallbackGraph$.
Specifically,
when the analysis needs to propagate
the resulting state from the exit point of a callback $x$,
instead of propagating that state to the caller
(note that the caller of a callback is the event loop),
it propagates it to the entry points
of the next callbacks,
i.e., all callback nodes $y \in Node$
where $x \rightarrow_{cg} y$ holds.
In other words,
the edges of a callback graph
reflect how the state is propagated
from the exit point of a callback node $x$
to the entry point of a callback node $y$.
Obviously,
if there is not any path between two nodes
in the graph,
that is,
$x \not\rightarrow_{cg}^{*} y$,
and $y \not\rightarrow_{cg}^{*} x$,
we propagate the state coming from the exit point of
$x$ to the entry point of $y$
and vice versa.

\point{Remark}
Callback-sensitivity does not work with contexts
to improve the precision of the analysis.
Therefore,
we still represent the event loop
as a single program point.
As a result,
the state produced
by the last executed callbacks is propagated to the event loop,
leading to the join of this state
with the initial one.
The join of those states
is then again propagated across the nodes of
the callback graph until convergence.
Therefore,
there is still some imprecision.
However,
callback-sensitivity minimizes
the number of those joins,
as they are only caused by the callbacks invoked last.

\subsubsection{Context-Sensitivity}
\label{sec:context-sen}

\begin{figure}
\begin{lstlisting}
function foo() { ... }

var x = Promise.resolve()
  .then(foo)
  .then(function ff1() { ... })
  .then(foo)
  .then(function ff2() { ... })
  .then(foo)
  .then(function ff3() { ...  });
\end{lstlisting}
\caption{An example program where we create a promise chain.
Notice that we register the function {\tt foo} multiple times across the chain.}\label{fig:ex-context}
\end{figure}

\begin{figure}[t]
    \centering
    \subcaptionbox{QR-insensitive analysis\label{fig:context-cg1}}{\includegraphics[width=0.29\linewidth]{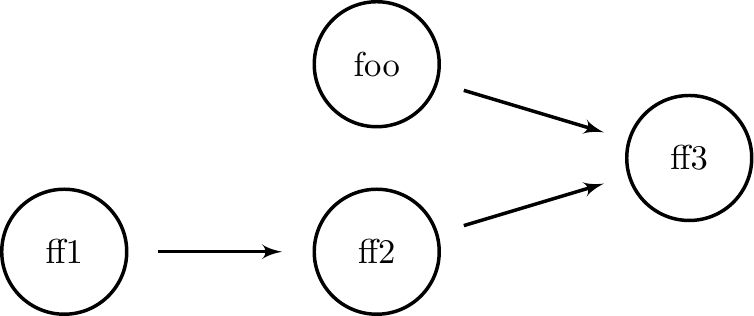}}%
    \hfill
    \subcaptionbox{QR-sensitive analysis\label{fig:context-cg2}}{\includegraphics[width=0.7\linewidth]{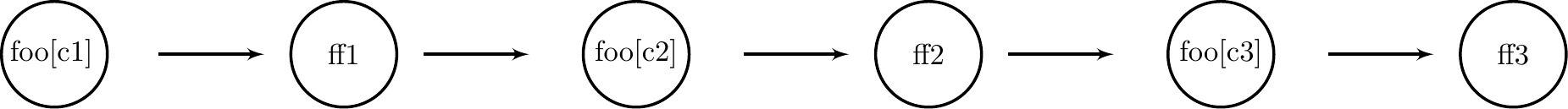}}%
\caption{Callback graph of program of Figure~\ref{fig:ex-context}
    produced by the QR-insensitive and QR-sensitive analysis respectively.}\label{fig:context-cg}
\end{figure}

Recall from Section~\ref{sec:callback-graph}
that a callback graph is defined
as $\mathcal{P}(Node \times Node)$,
where $n \in Node = C \times F$.
It is possible to increase the precision of a callback graph
(and therefore the precision of the analysis)
by distinguishing callbacks based on
the context in which they are invoked.
Existing flavors of context-sensitivity are not
so useful in differentiating asynchronous functions from each other.
For instance,
object-sensitivity~\cite{object-sen,context-sen},
which separates invocations based on
the value of the receiver---and
has been proven to be particularly effective
for the analysis of object-oriented languages---is
not fruitful in the context of
asynchronous callbacks
because in most cases
the receiver of callbacks corresponds to the global object.
Similarly,
previous work in the static analysis of JavaScript~\cite{tajs,jsai}
creates a context with regards to the arguments of a function.
Such a strategy might not be potent in cases
where a callback expects no arguments
or the arguments from two different calls are indistinguishable.

We introduce one novel context-sensitivity flavor---which
we call \emph{QR-sensitivity}---as
an effort to boost the analysis precision.
QR-sensitivity separates callbacks according to
1) the queue object to which they are added (Q),
and 2) the queue object their return value fulfills (R).
In this case,
the domain of contexts is given by:

$$c \in C = \widehat{Addr} \times \widehat{Addr}$$

In other words,
every context is
a pair $(l_q, l_r) \in \widehat{Addr} \times \widehat{Addr}$,
where $l_q$ stands for the allocation site
of the queue object to which we add a callback,
and $l_r$ is the abstract address of the queue object
which the return value of a callback fulfills.
Notice that this domain is finite;
thus,
the analysis always terminates.
\\

\noindent
{\bf Example:} As a motivating example,
consider the program of Figure~\ref{fig:ex-context}.
This program creates a promise chain
where we register different callbacks
at every step of the asynchronous computation.
At line 1,
we define the function {\tt foo()}.
We asynchronously call {\tt foo()}
multiple times,
i.e., at lines 4, 6, and 8.
Recall that the chains of promises
enable us to enforce a deterministic execution
of the corresponding callbacks.
Specifically,
based on the actual execution,
the event loop invokes the callbacks in the following order:
${\tt foo()} -> {\tt ff1()} -> {\tt foo()} -> {\tt ff2()} -> {\tt foo()} -> {\tt ff3()}$.
Figure~\ref{fig:context-cg1}
presents the callback graph
of the program of our example
produced by a QR-insensitive analysis.
In this case,
the analysis considers the different invocations of {\tt foo()}
as identical.
As a result,
the analysis loses the temporal relation between
{\tt foo()}
and {\tt ff1()}, {\tt ff2()}---indicated
by the fact that the respective nodes are not connected
to each other---because {\tt foo()}
is also called both before and after
{\tt ff1()} and {\tt ff2()}.
On the contrary,
a QR-sensitive analysis ends up
with an entirely precise callback graph
as shown in Figure~\ref{fig:context-cg2}.
The QR-sensitive analysis distinguishes
the different invocations of {\tt foo()} from each other
because it creates three different contexts;
one for every call of {\tt foo()}.
Specifically,
we have $c_1=(l_3,l_4), c_2=(l_5,l_6), c_3=(l_7, l_8)$,
where $l_i$ stands for the promise object allocated at line $i$.
For example,
the second invocation of {\tt foo()}
is related to the promise object
created by the call of {\tt then()} at line 5,
and its return value fulfills the promise object
allocated by the invocation of {\tt then()} at line 6.

\subsection{Implementation}

\begin{figure}
\begin{lstlisting}
function open(filename, flags, mode, callback) {
    TAJS_makeContextSensitive(open, 3);
    var err = TAJS_join(TAJS_make("Undef"), TAJS_makeGenericError());
    var fd = TAJS_join(TAJS_make("Undef"), TAJS_make("AnyNum"));
    TAJS_addAsyncIOCallback(callback, err, fd);
}

var fs = {
  open: open
  ...
}
\end{lstlisting}
\caption{{A model for {\tt fs.open} function.}
All functions starting with {\tt TAJS\_}
are special functions whose
body does not correspond to any node in the CFG.
They are just hooks for producing side-effects to the state
or evaluating to some value,
and their models are implemented in Java.
For instance,
{\tt TAJS\_make("AnyStr")} evaluates to a value
that can be any string.}\label{fig:ex-model}
\vspace{-4mm}
\end{figure}

Our prototype implementation extends~\emph{TAJS}~\cite{tajs,lazypropagation,jensendom};
a state-of-the-art static analyzer for JavaScript.
TAJS analysis is implemented
as an instance of the abstract interpretation framework~\cite{abstract-inter},
and it is designed to be sound.
It uses a lattice specifically designed for JavaScript
which is capable of handling
the vast majority of JavaScript's complicated features
and semantics.
TAJS analysis is both flow- and context-sensitive.
The output of the analysis is the set of all
reachable states from an initial state
along with a call graph.
TAJS can detect
various type-related errors such as
the use of a non-function variable in a call expression,
property access of {\tt null} or {\tt undefined} variables,
inconsistencies caused by implicit type conversions,
and many others~\cite{tajs}.

Prior to our extensions,
TAJS consisted of approximately \nnum{83500} lines of Java code.
The size of our additions is roughly \nnum{6000} lines of Java code.
Our implementation is straightforward
and is guided by the design of our analysis.
Specifically,
we first incorporate the domains
presented in Figure~\ref{fig:tajs-async-domains}
into the definition of the abstract state of TAJS.
Then,
we provide models for promises
written in Java by faithfully
following the ECMAScript specification.
Recall again that
our models exploit the $\lambda_{\tt q}$ calculus
presented in Section~\ref{sec:modeling}
and they produce side-effects
that over-approximate the behavior of JavaScript promises.
Beyond that,
we implement models for the special constructs of $\lambda_{\tt q}$,
i.e., {\tt addTimerCallback}, {\tt addIOCallback}
(Recall Section~\ref{sec:model-timers-io}),
which are used for adding callbacks
to the timer- and asynchronous I/O-related queue objects
respectively.
We implement the models for timers in Java;
however,
we write JavaScript models for asynchronous I/O operations,
when it is necessary.

For example,
Figure~\ref{fig:ex-model}
shows the JavaScript code that models
function {\tt open()} of the Node.js module {\tt fs}.
In particular,
{\tt open()} asynchronously opens a given file.
When I/O operation completes,
the callback provided by the developer is called
with two arguments:
1) {\tt err} which is not {\tt undefined}
if there is an error during I/O,
2) {\tt fd} which is an integer indicating
the file descriptor of the opened file.
Note that
{\tt fd} is {\tt undefined},
if any error occurs.
Our model first makes
{\tt open()} parameter-sensitive
on the third argument
which corresponds to the callback provided by the programmer.
Then,
at lines 3 and 4,
it initializes the arguments of the callback,
i.e., {\tt err} and
{\tt fd} respectively.
Observe that we initialize those arguments
so that they capture
all the possible execution scenarios,
i.e., {\tt err} might be {\tt undefined} or
point to an error object,
and {\tt fd} might be {\tt undefined}
or any integer reflecting all possible file descriptors.
Finally,
at line 5,
we call the special function {\tt TAJS_addAsyncIOCallback()},
which registers the given callback on
the queue object responsible for I/O operations,
implementing the semantics of the {\tt addIOCallback} primitive
from our $\lambda_{\tt q}$ calculus.

\section{Empirical Evaluation}
\label{sec:evaluation}

In this section,
we evaluate our static analysis
on a set of hand-written micro-benchmarks
and a set of real-world JavaScript modules.
Then,
we experiment with different parameterizations of the analysis, 
and report the precision and performance metrics.

\subsection{Experimental Setup}
\begin{table}[t]
\centering
\resizebox{\linewidth}{!}{%
\begin{tabular}{|l|r|r|r|r|r|r|}
\hline
{\bf Benchmark}        & {\bf LOC}  & {\bf ELOC} & {\bf Files} & {\bf Dependencies} & {\bf Promises} & {\bf Timers/Async I/O}  \bigstrut\\
\hline\hline
\href{https://github.com/vitalets/controlled-promise}{controlled-promise} & 225  & 225 & 1 & 0 &  4 & 1 \\
\href{https://github.com/github/fetch}{fetch}  & 517  & \nnum{1118} & 1 & 1 & 26  & 2 \\
\href{https://github.com/kokororin/honoka}{honoka} & 324 & \nnum{1643} & 6 & 6 & 4 & 1 \\ 
\href{https://github.com/axios/axios}{axios}  & \nnum{1733} & \nnum{1733} & 26  & 0 & 7 & 2 \\ 
\href{https://github.com/alphasp/pixiv-api-client}{pixiv-client} & \nnum{1031} & \nnum{3469} & 1 & 2 & 64 & 2 \\
\href{https://github.com/isaacs/node-glob}{node-glob} & \nnum{1519} & \nnum{6131} & 3 & 6 & 0 & 5 \\
\hline
\end{tabular}
}
\caption{{List of selected macro-benchmarks
  and their description.
  Each benchmark is described by
  its lines of code (LOC),
  its lines of code including its dependencies (ELOC),
  number of files,
  number of dependencies,
  number of promise-related statements
  (e.g., {\tt Promise.resolve()},
  {\tt Promise.reject()},
  {\tt then()}, etc.),
  and number of statements associated with timers
  (e.g., {\tt setTimeout()}, {\tt setImmediate()}, etc.) or
  asynchronous I/O
  (e.g., asynchronous file system or network operations etc.).
}}
\label{tb:benchmarks}
\vspace{-4mm}
\end{table}

To test that our technique behaves as expected
we first wrote a number of micro-benchmarks.
Each of those programs consists of approximately 20--50 lines of code
and examines certain parts of the analysis.
Beyond micro-benchmarks,
we evaluate our analysis on 6 real-world JavaScript modules.
The most common macro benchmarks
for static analyses used in the literature
are those provided by
JetStream\footnote{\url{https://browserbench.org/JetStream/}},
and V8 engine\footnote{\url{http://www.netchain.com/Tools/v8/}}\cite{tajs,jsai,safe-tunable}.
However,
those benchmarks are not asynchronous;
thus,
they are not suitable for evaluating our analysis.
To find interesting benchmarks,
we developed an automatic mechanism for
collecting and analyzing Github repositories.
First,
we collected a large number of Github repositories
using two different options.
The first option extracted the
Github repositories of the most depended upon
{\tt npm} packages\footnote{\url{https://www.npmjs.com/browse/depended}}.
The second option employed
the Github API\footnote{\url{https://developer.github.com/v3/}}
to find JavaScript repositories
which are related to promises.
We then investigated the Github repositories
which we collected at the first phase
by computing various metrics
such as lines of code,
number of promise-, timer-
and asynchronous IO-related statements.
We manually selected the 6 JavaScript modules
presented in Table~\ref{tb:benchmarks}.
Most of them are libraries
for performing HTTP requests
or file system operations.

We experiment with 4 different analyses:
1) an analysis which is neither callback- nor QR-sensitive (NC-No),
2) a callback-insensitive but QR-sensitive analysis (NC-QR),
3) a callback-sensitive but QR-insensitive analysis (C-No),
and 4) a both callback- and QR-sensitive analysis (C-QR).
We evaluate the precision of each analysis
in terms of the number of the analyzed callbacks,
the precision of the computed callback graph,
and the number of reported type errors.
We define
the precision of a callback graph as
the quotient between
the number of callback pairs
whose execution order is determined
and the total number of callback pairs.
Also,
we embrace a client-based precision metric,
i.e.,
the number of reported type errors
as in the work of~\cite{jsai}.
The fewer type errors an analysis reports,
the more precise it is.
The same applies to the number of callbacks
inspected by the analysis;
fewer callbacks indicate a more accurate analysis.
To compute the performance characteristics of every analysis,
we re-run every experiment ten times
in order to receive reliable measurements.
All the experiments were run on
a machine with an Intel i7 2.4{\tt GHz} quad-core processor
and 8{\tt GB} of RAM.

\begin{table}[t!]
\centering
\resizebox{\textwidth}{!}{%
\begin{tabular}{|l|r|r|r|r||r|r|r|r||r|r|r|r|}
\hline
           & \multicolumn{4}{c||}{\textbf{Analyzed Callbacks}} & \multicolumn{4}{c||}{\textbf{Callback Graph Precision}} & \multicolumn{4}{c|}{\textbf{Type Errors}}     \bigstrut\\
\hline\hline
  {\bf Benchmark} & {\bf NC-No} & {\bf NC-QR} & {\bf C-No}  & {\bf C-QR}  & {\bf NC-No} & {\bf NC-QR} & {\bf C-No} & {\bf C-QR} & {\bf NC-No} & {\bf NC-QR} & {\bf C-No}  & {\bf C-QR} \\
\hline
micro01   & 5  &  5  & 4  & 4  &0.8  &  0.8  & 1.0  & 1.0  &2   & 2  & 0 & 0\\
micro02   & 3  &  3  & 3  & 3  &1.0  &  1.0  & 1.0  & 1.0  &1   & 1  & 0 & 0\\
micro03   & 2  &  2  & 2  & 2  &1.0  &  1.0  & 1.0  & 1.0  &1   & 1  & 0 & 0\\
micro04   & 4  &  4  & 4  & 4  &0.5  &  0.5  & 0.5  & 0.5  &1   & 1  & 1 & 1\\
micro05   & 8  &  8  & 7  & 7  &0.96 &  0.96 & 1.0  & 1.0  &3   & 3  & 0 & 0\\
micro06   & 11 &  11 & 11 & 11 &1.0  &  1.0  & 1.0  & 1.0  &3   & 3  & 1 & 1\\
micro07   & 14 &  14 & 13 & 13 &0.86 &  0.87 & 1.0  & 1.0  &1   & 1  & 0 & 0\\
micro08   & 5  &  5  & 5  & 5  &0.8  &  0.8  & 0.8  & 0.8  &1   & 1  & 0 & 0\\
micro09   & 5  &  5  & 4  & 4  &0.9  &  0.9  & 1.0  & 1.0  &1   & 1  & 0 & 0\\
micro10   & 3  &  3  & 3  & 3  &1.0  &  1.0  & 1.0  & 1.0  &1   & 1  & 1 & 1\\
micro11   & 4  &  4  & 4  & 4  &0.83 &  0.83 & 0.83 & 0.83 &5   & 5  & 5 & 5\\
micro12   & 5  &  5  & 5  & 5  &0.9  &  0.9  & 1.0  & 1.0  &2   & 2  & 0 & 0\\
micro13   & 4  &  4  & 3  & 3  &0.83 &  0.83 & 1.0  & 1.0  &1   & 1  & 0 & 0\\
micro14   & 6  &  6  & 5  & 5  &0.8  &  0.8  & 1.0  & 1.0  &2   & 2  & 0 & 0\\
micro15   & 6  &  6  & 6  & 6  &0.8  &  0.8  & 1.0  & 1.0  &0   & 0  & 0 & 0\\
micro16   & 6  &  6  & 6  & 6  &1.0  &  1.0  & 1.0  & 1.0  &1   & 1  & 0 & 0\\
micro17   & 3  &  3  & 3  & 3  &0.67 &  0.67 & 0.67 & 0.67 &2   & 2  & 2 & 2\\
micro18   & 4  &  3  & 4  & 3  &0.83 &  1.0  & 0.83 & 1.0  &1   & 0  & 1 & 0\\
micro19   & 14 &  7  & 14  & 7 &0.73 &  0.93 & 0.74 & 1.0  &0   & 0  & 0 & 0\\
micro20   & 6  &  6  & 6  & 6  &0.93 &  0.93 & 1.0  & 1.0  &0   & 0  & 0 & 0\\
micro21   & 5  &  5  & 4  & 4  &0.9  &  0.9  & 1.0  & 1.0  &1   & 1  & 0 & 0\\
micro22   & 6  &  6  & 5  & 5  &0.87 &  0.87 & 0.9  & 0.9  &1   & 1  & 0 & 0\\
micro23   & 6  &  6  & 5  & 5  &0.87 &  0.87 & 1.0  & 1.0  &3   & 3  & 1 & 1\\
micro24   & 3  &  3  & 3  & 3  &1.0  &  1.0  & 1.0  & 1.0  &2   & 2  & 1 & 1\\
micro25   & 8  &  8  & 8  & 8  &0.79 &  0.79 & 0.79 & 0.79 &1   & 1  & 0 & 0\\
micro26   & 9  &  9  & 7  & 7  &0.89 &  0.89 & 1.0  & 1.0  &3   & 3  & 1 & 1\\
micro27   & 3  &  3  & 3  & 3  &1.0  &  1.0  & 1.0  & 1.0  &1   & 1  & 1 & 1\\
micro28   & 7  &  7  & 7  & 7  &0.81 &  0.81 & 0.81 & 0.81 &1   & 1  & 1 & 1\\
micro29   & 4  &  4  & 4  & 4  &0.5  &  1.0  & 0.5  & 1.0  &0   & 0  & 0 & 0\\
\hline
  {\bf Average} & {\bf 5.83} & {\bf 5.55} & {\bf 5.45} & {\bf 5.17} & {\bf 0.85} & {\bf 0.88} & {\bf 0.91} & {\bf 0.94}  & {\bf 1.45} & {\bf 1.41} & {\bf 0.55} & {\bf 0.52} \bigstrut\\
\hline
  {\bf Total} & {\bf 169} & {\bf 161} & {\bf 158} & {\bf 150} & & & & & {\bf 42} & {\bf 41} & {\bf 16} & {\bf 15} \bigstrut \\
\hline
\end{tabular}}
\caption{{Precision on micro-benchmarks.}}
\label{table:micro-precision}
\vspace{-4mm}
\end{table}

\subsection{Results}

{\bf Micro-benchmarks.} Table~\ref{table:micro-precision} shows
how precise every analysis is on every micro-benchmark.
Starting with callback-insensitive analyses
(i.e., columns NC-No and NC-QR),
we observe that in general QR-sensitivity improves
the precision of the callback graph by~\empirical{3.6\%} on average.
That small boost of QR-sensitivity
is explained by the fact that \emph{only}~\empirical{3} out of~\empirical{29} micro-benchmarks invoke the same callback multiple times.

Recall from Section~\ref{sec:context-sen}
that QR-sensitivity is used to
distinguish different calls of the same callback.
Therefore,
if one program does not use a specific callback multiple times,
QR-sensitivity does not make any difference.
However, if we focus on the results of the micro-benchmarks
where we come across such behaviors,
i.e. micro18,
micro19,
and micro29,
we get a significant divergence
of the precision of callback graph.
Specifically,
QR-sensitivity improves precision by~\empirical{20.5\%},
\empirical{27.4\%}
and~\empirical{100\%} in
micro18,
micro19
and micro29 respectively.
Besides that,
in micro19,
there is a striking decrease in the number of the analyzed callbacks:
the QR-insensitive analyses inspect~\empirical{14} callbacks
compared to the QR-sensitive analyses
which examine only~\empirical{7}.

The results regarding the number of type errors
are almost identical for every analysis:
a QR-insensitive analysis
reports~\empirical{42} type errors in total,
whereas all the other QR-sensitive analyses
produce warnings for~\empirical{41} cases.

Moving to callback-sensitive analyses, the results indicate clear differences. 
First, a callback-sensitive but QR-insensitive analysis
reports only~\empirical{16} type errors in total
(i.e., \empirical{61.9\%} fewer type errors than callback-insensitive analyses), and amplifies the average precision of the callback graph
from~\empirical{0.85} to~\empirical{0.91}.
As before,
the QR-sensitive analyses boost the precision of the callback graph
by~\empirical{20.4\%}, 
\empirical{35.1\%},
and~\empirical{100\%} in micro18,
micro19, and micro29 respectively.
Finally,
a callback-sensitive
and QR-insensitive analysis decreases
the total number of the analyzed callbacks
from~\empirical{169} to~\empirical{158}.
Notice that if callback-sensitivity
and QR-sensitivity are combined,
the total number of callbacks
is reduced by~\empirical{11.2\%}.
\\


\begin{table}[tp]
\centering
\resizebox{\textwidth}{!}{%
\begin{tabular}{|l|r|r|r|r||r|r|r|r||r|r|r|r|}
\hline
           & \multicolumn{4}{c||}{\textbf{Analyzed Callbacks}} & \multicolumn{4}{c||}{\textbf{Callback Graph Precision}} & \multicolumn{4}{c|}{\textbf{Type Errors}}     \bigstrut\\
\hline\hline
  {\bf Benchmark} & {\bf NC-No} & {\bf NC-QR} & {\bf C-No}  & {\bf C-QR}  & {\bf NC-No} & {\bf NC-QR} & {\bf C-No} & {\bf C-QR} & {\bf NC-No} & {\bf NC-QR} & {\bf C-No}  & {\bf C-QR} \\
\hline
controlled-promise & 6  & 6  & 6  & 6  & 0.866 & 0.905 & 0.866 & 0.905 & 3  & 3  & 2  & 2  \\
fetch              & 22 & 22 & 19 & 19 & 0.829 & 0.956 & 0.822 & 0.972 & 8  & 8  & 6  & 6  \\
honoka             & 8  & 8  & 6  & 6  & 0.929 & 0.929 & 1.0   & 1.0   & 1  & 1  & 0  & 0  \\
axios              & 15 & 15 & 14 & 14 & 0.678 & 0.83  & 0.686 & 0.871 & 2  & 2  & 1  & 1  \\
pixiv-client       & 18 & 18 & 17 & 15 & 0.771 & 0.803 & 0.794 & 0.863 & 3  & 3  & 3  & 2 \\
node-glob          & 3  & 3  & 3  & 3  & 0.667 & 0.667 & 0.667 & 0.667 & 19 & 19 & 19 & 19 \\
\hline
  {\bf Average} & {\bf 12} & {\bf 12} & {\bf 10.8} & {\bf 10.5}  & {\bf 0.79} & {\bf 0.848} & {\bf 0.805} & {\bf 0.88}  & {\bf  6} & {\bf 6} & {\bf 5.1} & {\bf 5} \bigstrut\\
\hline
  {\bf Total} & {\bf 72} & {\bf 72} & {\bf 65} & {\bf 63} & & & & & {\bf 36} & {\bf 36} & {\bf 31} & {\bf 30}\bigstrut \\ 
\hline
\end{tabular}}
\caption{{Precision on macro-benchmarks.}}
\label{table:macro-precision}
\vspace{-4mm}
\end{table}

\noindent
{\bf Macro-benchmarks.}
Table~\ref{table:macro-precision} reports
the precision metrics of every analysis of the macro-benchmarks.
First, we make similar observations as those of micro-benchmarks.
In general,
QR-sensitivity leads to a more precise callback graph
for~4 out of~6 benchmarks.
The improvement ranges from~\empirical{4.6\%}
to~\empirical{26.9}\%. 
On the other hand, callback-sensitive analyses contribute to fewer type errors for~5 out of~6 benchmarks
reporting~\empirical{16.7\%} fewer type errors in total.
Additionally,
if we combine QR- and callback-sensitivity,
we can boost the analysis precision for~5 out of~6 benchmarks.
Specifically,
the QR- and callback-sensitive analysis
improves the callback graph precision by up to~\empirical{28.5\%}
(see the {\tt axios} benchmark),
and achieves a~\empirical{88\%}
callback graph precision on average.
On the other hand,
the naive analysis (neither QR- nor callback-sensitive)
reports only a~\empirical{79\%} precision for callback graph
on average.

By examining the results for the {\tt node-glob} benchmark,
we see that every analysis produces identical results.
{\tt node-glob} uses only timers
and asynchronous I/O operations.
Neither callback- nor QR-sensitivity is effective
for that kind of benchmarks,
since we follow a conservative approach
for modeling the execution order of timers and asynchronous I/O,
regardless of the registration order of their callbacks.
For example,
we assume that two callbacks $x$ and $y$
are executed in any order,
even if $x$ is scheduled before $y$ (and vice versa).
Therefore, keeping a more precise state does not lead to
a more precise callback graph.

\begin{table}[tb]
\centering
\resizebox{\textwidth}{!}{%
\begin{tabular}{|l|r|r|r|r||r|r|r|r|}
\hline
         & \multicolumn{4}{c||}{\textbf{Average Time}}      & \multicolumn{4}{c|}{\textbf{Median}} \bigstrut\\
\hline\hline
  {\bf Benchmark} & {\bf NC-No} & {\bf NC-QR}  & {\bf C-No} & {\bf C-QR} & {\bf NC-No} & {\bf NC-QR}  & {\bf C-No} & {\bf C-QR}\bigstrut\\
\hline
controlled-promise & 2.3   & 2.22  & 2.27  & 2.28  & 2.29  & 2.26  & 2.25  & 2.31  \\
fetch              & 8.53  & 7.97  & 7.07  & 6.98  & 8.52  & 8.26  & 7.46  & 7.22  \\
honoka             & 4.14  & 4.05  & 3.86  & 3.94  & 4.12  & 4.0   & 3.61  & 3.81  \\
axios              & 6.99  & 7.86  & 6.74  & 8.32  & 7.02  & 8.0   & 6.94  & 8.37  \\
pixiv-client       & 22.11 & 24.92 & 24.77 & 28.89 & 22.19 & 25.16 & 24.65 & 29.2  \\
node-glob          & 15.55 & 16.71 & 15.46 & 14.47 & 16.62 & 16.71 & 16.17 & 15.74 \\
\hline
\end{tabular}}
\caption{{Times of different analyses in seconds.}}
\label{table:macro-performance}
\vspace{-6mm}
\end{table}

Table~\ref{table:macro-performance}
gives the running times of every analysis
on macro-benchmarks.
We notice that
in some benchmarks
(such as fetch)
a more precise analysis may decrease the running times
by~\empirical{3\%--18\%}.
This is justified
by the fact that a more precise analysis
might compress the state faster
than an imprecise analysis.
For instance,
in fetch,
an imprecise analysis led to the analysis
of 3 spurious callbacks,
yielding to a higher analysis time.
The results appear to be consistent
with those of the recent literature
which suggests that
precision might lead to a faster analysis
in some cases~\cite{parkdom}.
On the other hand,
we observe a non-trivial decrease in the analysis performance
in only benchmark.
Specifically,
the analysis sensitivity increased
the running times of {\tt pixiv-client}
by~\empirical{12\%--30.6\%}.
However,
such an increase seems to be acceptable.

\subsection{Case Studies}

\begin{figure}[t!]
\begin{multicols}{2}
\begin{lstlisting}[basicstyle=\scriptsize\ttfamily]
function consumed(body) {
    if (body.bodyUsed) {
        return Promise.reject(new TypeError("Already read"));
    }
    body.bodyUsed = true;
}
...
function Body() {
  ...
  this.bodyUsed = false;
  this._bodyInit = function() {
    ...
    if (typeof body === "string") {
          this._bodyText = body;
    } else if (Blob.prototype.isPrototypeOf(body)) {
        this._bodyBlob = body;
    }
    ...
  }
  this.text = function text() {
        var rejected = consumed(this);
        if (rejected) {
            return rejected;
        }
        if (this._bodyBlob) {
            return readBlobAsText(this._bodyBlob);
        } else if (this._bodyArrayBuffer) {
            return Promise.resolve(readArrayBufferAsText(this._bodyArrayBuffer));
        } else if (this._bodyFormData) {
            throw new Error("could not read FormData body as text");
        } else {
            return Promise.resolve(this._bodyText);
        }
    };
  ...
  this.formData = function formData() {
       return this.text().then(decode);
  }
}
...
function Response(body) {
  ...
  this._bodyInit(body);
}
Body.call(Response.prototype);
...
function fetch(input, init) {
  return new Promise(function (resolve, reject) {
    ...
    var xhr = new XMLHttpRequest();
    xhr.onload = function onLoad() {
      ...
      resolve(new Response(xhr.response));
    }
  });
}
\end{lstlisting}
\end{multicols}
\caption{Code fragment taken from fetch.}\label{fig:fetch1}
\end{figure}

In this section,
we describe some case studies
coming from the macro-benchmarks.

{\bf fetch.} Figure~\ref{fig:fetch1} shows a code fragment
taken from fetch\footnote{\url{https://github.com/github/fetch}}.
Note that we omit irrelevant code for brevity.
The function {\tt Body()} defines a couple of methods
(e.g., {\tt text()}, {\tt formData()})
for manipulating the body of a response.
Observe that those methods are registered on
the prototype of {\tt Response} using
the function {\tt Function.prototype.call()} at line 45.
Note that {\tt Body} also contains a method
(i.e., {\tt \_initBody()})
for initializing the body of
a response according to the type of the input.
To this end,
the {\tt Response} constructor takes a body
as a parameter and initializes it
through the invocation of {\tt \_initBody()} (lines 41, 43).
The function {\tt text()}
reads the body of a response in a text format
(lines 20--34).
If the body of the response has been already read,
{\tt text()} returns a rejected promise (lines 3, 22--23).
Otherwise,
it marks the property {\tt bodyUsed}
of the response object as true
(line 5),
and then it returns
a fulfilled promise depending on
the type of the body of the provided response (lines 25--33).
The function {\tt formData()} (lines 36--38) asynchronously reads
the body of a response in a text format,
and then it parses it into a {\tt FormData} object\footnote{\url{https://developer.mozilla.org/en-US/docs/Web/API/FormData}}
through the call of the function {\tt decode()}.
The function {\tt fetch()} (lines 47--56) makes
a new asynchronous request.
When the request completes successfully,
the callback {\tt onLoad()} is executed asynchronously
(line 51).
That callbacks finally fulfills
the promise returned by {\tt fetch()}
with a new response object initialized with the response of
the server (line 53).

\begin{figure}[t]
\begin{adjustbox}{minipage=\linewidth,margin=20pt \smallskipamount,center}
  \centering
  \begin{minipage}{.5\linewidth}
  \begin{lstlisting}[caption={{Case 1}}, label={lst:fetch-case1},
                     basicstyle=\scriptsize\ttfamily]
    fetch("/helloWorld").then(function foo(value) {
      var formData = value.formData();
      // Do something with form data.
    })
  \end{lstlisting}
  \end{minipage}\hfill
  \begin{minipage}{.5\linewidth}
  \begin{lstlisting}[caption={{Case 2}}, label={lst:fetch-case2},
                     basicstyle=\scriptsize\ttfamily]
    var response = new Response("foo=bar");
    var formData = response.formData();
    var response2 = new Response(new Blob("foo=bar"));
    var formData2 = response2.formData();
  \end{lstlisting}
  \end{minipage}
\caption{{Code fragments which use the {\tt fetch} API.}}\label{fig:fetch-cases}
\end{adjustbox}
\end{figure}

In Listing~\ref{lst:fetch-case1},
we make an asynchronous request to the endpoint ``/helloWorld''
using the {\tt fetch} API.
Upon success,
we schedule the callback {\tt foo()}.
Recall that the parameter {\tt value} of {\tt foo()}
corresponds to the response object coming from line 53
(Figure~\ref{fig:fetch1}).
In {\tt foo()},
we convert the response of the server
into a {\tt FormData} object (line 2).
A callback-insensitive analysis,
which considers that
the event loop executes all callbacks in any order,
merges all the data flow stemming from those callbacks into a single point.
As a result,
the side effects of {\tt onLoad()}
and {\tt foo()} are directly propagated
to the event loop.
In turn,
the event loop propagates the resulting state again
to those callbacks.
This is repeated until convergence.
Specifically,
the callback {\tt foo()} calls {\tt value.formData()},
which updates the property {\tt bodyUsed} of
the response object to true
(Figure~\ref{fig:fetch1}, line 5).
The resulting state is propagated to the event loop
where is joined with the state
which stems from the callback {\tt onLoad()}.
Notice that the state of {\tt onLoad()}
indicates that {\tt bodyUsed} is false
because the callback {\tt onLoad()} creates a fresh response object.
(Figure~\ref{fig:fetch1}, lines 10, 53).
The join of those states
changes the abstract value of {\tt bodyUsed}
to $\top$.
That change is propagated again to {\tt foo()}.

This imprecision makes
the analysis to consider both {\tt if}
and {\tt else} branches at lines 2--5.
Thus,
the analysis allocates a rejected promise at line 3,
as it mistakenly considers
that the body has been already consumed.
This makes {\tt consumed()}
return a value
that is either {\tt undefined}
or a rejected promise at line 23.
The value returned by {\tt consumed()} is finally propagated to
{\tt formData()} at line 37,
where the analysis reports a false positive;
a property access of an {\tt undefined} variable
(access of the property ``then''),
because {\tt text()} might return an undefined variable
due to the return statement at line 26.
A callback-sensitive analysis neither
reports a type error at line 43
nor creates a rejected promise at line 4.
It respects the execution order of callbacks,
that is,
the callback {\tt foo()} is executed \emph{after}
the callback {\tt onLoad()}.
Therefore,
the analysis propagates
a more precise state to the entry of {\tt foo()}:
the state resulted
by the execution of {\tt onLoad()},
where a new response object is initialized
with the field {\tt bodyUsed} set to false.

In Listing~\ref{lst:fetch-case2},
we initialize a response object
with a body which has a string type (line 2).
In turn,
by calling the {\tt formData()} method,
we first read the body of the response in a text format,
and then we decode it into a {\tt FormData} object
by asynchronously calling the {\tt decode()} function
(Figure~\ref{fig:fetch1}, line 37).
Since the body of the response is already in a text format,
{\tt text()} returns a fulfilled promise
(Figure~\ref{fig:fetch1}, line 32).
At the same time,
at line 5 of Listing~\ref{lst:fetch-case2},
we allocate a fresh response object
whose body is an instance of {\tt Blob}\footnote{\url{https://developer.mozilla.org/en-US/docs/Web/API/Blob}}.
Therefore,
calling {\tt formData()} schedules function {\tt decode()} again.
However
this time,
the callback {\tt decode()} is registered on a different promise
because the second call of {\tt text()} returns a promise
created by the function {\tt readBlobAsText()}
(Figure~\ref{fig:fetch1}, line 26).
A QR-sensitive analysis---which creates
a context according to the queue object
on which a callback is registered---is
capable of separating the two invocations of {\tt decode()}
because the first call of {\tt decode()} is registered
on the promise object
which comes from line 32,
whereas the second call of {\tt decode()} is added
to the promise created by {\tt readBlobAsText()} at line 26.

{\bf honoka.} We return back to Figure~\ref{fig:motiv-ex-honoka}.
Recall that a callback-insensitive analysis
reports a spurious type error at line 17
when we try to access the property {\tt headers}
of {\tt honoka.response}
because it considers the case
where the callback defined at lines 15--23
is executed before that defined at lines 2--14.
Thus,
{\tt honoka.response} might be uninitialized
(recall that {\tt honoka.response} is initialized
during the execution of the first callback at line 3).
On the other hand,
a callback-sensitive analysis consults the callback graph
when it is time to propagate the state
from the exit point of a callback
to the entry point of another.
In particular,
when we analyze the exit node of the first callback,
we propagate the current state
to the second callback.
Therefore,
the entry point of the second function
has a state
which contains a precise value for {\tt honoka.response},
that is,
the object coming from the assignment at line 3.

\subsection{Threats to Validity}

Below we pinpoint the main threats to the validity of our results:
\begin{itemize}

\item Our analysis is an extension of an existing analyzer,
i.e., TAJS.
Therefore,
the precision and performance of TAJS
play an important role on
the results of our work.

\item Even though our analysis is designed to be sound,
it models some native functions of the JavaScript language unsoundly.
For instance,
we unsoundly model
the native function {\tt Object.freeze()},
which is used to prevent an object from being updated.
Specifically,
the model of {\tt Object.freeze()} simply
returns the object given as argument.

\item We provide manual models for some built-in Node.js modules
like {\tt fs}, {\tt http}, etc.\
or other APIs used in client-side applications
such as {\tt XMLHttpRequest}, {\tt Blob}, etc.
However,
manual modeling might neglect
some of the side-effects
which stem from the interaction with those APIs,
leading to unsoundness~\cite{gatekeeper,parkdom}.

\item Our macro-benchmarks consist of JavaScript libraries.
Therefore,
we needed to write some test cases
that invoke the API functions of those benchmarks.
We provided both hand-written test cases
and test cases or examples taken from
their documentation,
trying to test the main APIs
that exercise asynchrony in JavaScript.
\end{itemize}

\section{Related Work}
\label{sec:related}

In this section,
we briefly present previous work
related to the formalization
and program analysis for (asynchronous) JavaScript.

{\bf Semantics.}
Maffeis et al.~\cite{maffeis}
presented one of the first formalizations
of JavaScript by designing small-step operational semantics for
a subset of the 3rd version of ECMAScript.
In subsequent work,
Guha et al.\cite{ljs}
expressed the semantics of the 3rd edition of ECMAScript
through a different approach;
they developed a lambda calculus called $\lambda_{{\tt JS}}$,
and provided a desugaring mechanism for
converting JavaScript code into $\lambda_{{\tt JS}}$.
We used $\lambda_{{\tt JS}}$
as a base for modeling asynchronous JavaScript.
Later,
Gardner et al.~\cite{gardnerlogic}
introduced a program logic for reasoning about
client-side JavaScript programs
which support ECMAScript 3.
They presented big-step operational semantics
on the basis of that proposed by~\cite{maffeis},
and they introduced inference rules
for program reasoning
which are highly inspired from separation logic~\cite{separation}.
More recently,
Madsen et al.~\cite{promises}
and Loring et al.~\cite{asyncjs}
extended $\lambda_{{\tt JS}}$
for modeling promises
and asynchronous JavaScript respectively.
Our model is a variation of their works;
our modifications enable us to model
almost all the sources of asynchrony found in JavaScript---some of
them are not handled by their models.

{\bf Static Analysis for JavaScript.}
Guarnieri et al.~\cite{gatekeeper}
proposed one of the first pointer analyses
for a subset of JavaScript.
They precluded the use of {\tt eval}-family functions from
their analysis
as their work focused on widgets
where the use of {\tt eval} is not common.
It was one of the first works
that managed to model some of the peculiar features of JavaScript
such as prototype-based inheritance.
TAJS~\cite{tajs,lazypropagation,jensendom}
is a typer analyzer for JavaScript
which is implemented as a classical dataflow analysis.
Our work is implemented as
an extension of TAJS.
SAFE~\cite{safe} is
a static analysis framework,
which provides three different formal representations of JavaScript programs:
an abstract syntax tree (AST),
an intermediate language (IR)
and a control-flow graph (CFG).
SAFE implements a default analysis phase
which is plugged after the construction of CFG.
This analysis adopts a similar approach with that of TAJS,
i.e., a flow- and context-sensitive analysis
which operates on top of CFG.
JSAI~\cite{jsai} implements an analysis
through the abstract interpretation framework~\cite{abstract-inter}.
Specifically,
it employs a different approach
compared to other existing tools.
Unlike TAJS and SAFE,
JSAI operates on top of AST
rather than CFG;
it is flow-sensitive though.
To achieve this,
the abstract semantics is specified on
a CESK abstract machine~\cite{cesk},
which provides small-step reduction rules
and an explicit data structure (i.e., continuation)
which describes the rest of computation,
unwinding the flow of the program in this way.
The analysis is configurable with different flavors of
context-sensitivity
which are plugged into the analysis
through widening operator used in
the fix-point calculation~\cite{widening}.

Existing static analyses provide sufficient support
for precisely modeling browser environment.
Jensen et al.~\cite{jensendom} modeled HTML DOM
by creating a hierarchy of abstract states
which reflect the actual HTML object hierarchy.
Before the analysis begins,
an initial heap is constructed
which contains the set of the abstract objects
corresponding to the HTML code of the page.
Park et al.~\cite{parkdom} followed
a similar approach for modeling HTML DOM.
They also provided a more precise model
which respects the actual tree hierarchy of the DOM.
For example,
their model distinguishes
whether one DOM node
is nested to another or not.

{\bf Program Analysis for Asynchronous JavaScript Programs.}
The majority of static analyses for JavaScript treat
asynchronous programs conservatively~\cite{tajs,safe,jsai}---they assume
that the event loop processes all the asynchronous callbacks
in any order---leading to the analysis imprecision.
Also,
they focus on the client-side applications,
where asynchrony mainly appears in DOM events
and AJAX calls.
Madsen et al.~\cite{madsen2015ev} proposed
one of the first static analysis for server-side event-driven programs.
Although their approach is able to handle
asynchronous I/O operations---unlike our work---they
do not provide support for ES6 promises.
Additionally,
their work introduced a context-sensitivity strategy
which tries to imitate the different iterations of the event loop. 
However,
it imposes a large overhead on the analysis;
it is able to handle only small programs
(less than 400 lines of code).
In our work,
we propose callback-sensitivity
which improves precision without
highly sacrificing performance.
More recently,
Alimadadi et al.~\cite{promises2} presented
a dynamic analysis technique for detecting
promise-related errors
and anti-patterns in JavaScript programs.
Specifically, their approach exploits the promise graph;
a representation designed for debugging promise-based programs.
Beyond promises, our work also handles a broad spectrum of asynchronous features.

{\bf Race Detection.}
Zheng et al.~\cite{staticasync}
presented one of the first race detectors
by employing a static analysis for
identifying concurrency issues in asynchronous AJAX calls.
The aim of their analysis was to
detect data races between the code
which pre-processes an AJAX request
and the callback invoked when the response of the server is received.
A subsequent work~\cite{webracer}
adopted a dynamic analysis to detect data races in web applications.
They first proposed a happens-before relation model
to capture the execution order between different operations
that are present in a client-side application,
such as the loading of HTML elements, execution of scripts, etc.
Using this model,
their analyses reports data races,
by detecting memory conflicts between functions,
where there is not any happens-before relation to each other.
However, their approach introduced a lot of false positives
because most data races did not lead to severe concurrency bugs.
Mutlu et al.~\cite{datamatter} combined both
dynamic and static analysis and primarily focused on detecting data races
that have pernicious consequences on the correctness of applications,
such as those which affect the browser storage.
Initially, they collected the execution traces of an application,
and then, they applied a dataflow analysis on those traces
to identify data races.
Their approach effectively managed to report
a very small number of false positives.

\section{Conclusions \& Future Work}
\label{sec:conclusions}

Building upon previous works,
we presented the $\lambda_{\tt q}$ calculus
for modeling asynchrony in JavaScript.
Our calculus $\lambda_{\tt q}$ is flexible enough
so that we can express almost every asynchronous primitive
in the JavaScript language
up to the 7th edition of the ECMAScript.
We then presented an abstract version of $\lambda_{\tt q}$
which over-approximates the semantics of our calculus.

By exploiting that abstract version,
we designed and implemented what is,
to the best of our knowledge,
the first static analysis for dealing with
a wide range of asynchrony-related features.
At the same time,
we introduced the concept of callback graph;
a directed acyclic graph
which represents the temporal relations
between the execution of asynchronous callbacks,
and we proposed a more precise analysis,
i.e.~\emph{callback-sensitive} analysis
that respects the execution order of callbacks.
We parameterized our analysis
with a new context-sensitivity flavor
that is specifically used for asynchronous callbacks.

We then experimented
with different parameterizations of our analysis
on a set of hand-written
and real-world programs.
The results revealed
that we can analyze medium-sized JavaScript programs
using our approach.
The analysis sensitivity
(i.e., both callback- and context-sensitivity)
is able to ameliorate the analysis precision
without highly sacrificing performance.
Specifically,
as observed in the real-world modules,
our analysis achieves
a~\empirical{79\%} precision for the callback graph,
on average.
If we combine callback- and QR-sensitivity,
we can further improve the callback graph precision
by up to~\empirical{28.5\%}.
Also,
the callback- and QR-sensitive analysis
achieves a~\empirical{88\%} callback graph precision on average,
and reduces the total number of
type errors by~\empirical{16.7\%}.

Our work constitutes a general technique
that can be used as a base for further research.
Specifically, recent studies showed
that concurrency bugs found in JavaScript programs may sometimes be caused
by asynchrony~\cite{concstudy,fuzz}.
We could leverage our work to design a client analysis on top of it
so that it statically detects data races in
JavaScript programs.
Our callback graph might be an essential element
for such an analysis
because we can inspect it to identify
callbacks whose execution might be non-deterministic,
i.e., unconnected nodes in the callback graph.


\bibliography{jsan}

\end{document}